\documentclass[letterpaper,twocolumn,prl,aps,superscriptaddress,amsmath,amssymb,floatfix]{revtex4-1}
\usepackage{mathptmx}
\usepackage[latin9]{inputenc}
\usepackage{color}
\usepackage{amsmath}
\usepackage{amssymb}
\usepackage{graphicx}
\usepackage{esint}
\usepackage{enumitem}
\usepackage[unicode=true, bookmarks=true,bookmarksnumbered=false,bookmarksopen=false, breaklinks=false,pdfborder={0 0 1},backref=false,colorlinks=true]
 {hyperref}
\hypersetup{
 linkcolor=magenta,urlcolor=blue,citecolor=blue,pdfstartview={FitH}}

\makeatletter

\pdfpageheight\paperheight
\pdfpagewidth\paperwidth

\usepackage{textcomp}
\usepackage{epstopdf}

\pdfpageheight\paperheight
\pdfpagewidth\paperwidth

\@ifundefined{textcolor}{}{%
 \definecolor{BLACK}{gray}{0}
 \definecolor{WHITE}{gray}{1}
 \definecolor{RED}{rgb}{1,0,0}
 \definecolor{GREEN}{rgb}{0,1,0}
 \definecolor{BLUE}{rgb}{0,0,1}
 \definecolor{CYAN}{cmyk}{1,0,0,0}
 \definecolor{MAGENTA}{cmyk}{0,1,0,0}
 \definecolor{YELLOW}{cmyk}{0,0,1,0}
}

\usepackage{xcolor}
\usepackage{soul}
\definecolor{blue}{rgb}{0,0,1}
\definecolor{red}{rgb}{1,0,0}
\definecolor{green}{rgb}{0,1,0}

\usepackage{soul}

\makeatother

\begin{document}
\title{Suspension-Free Integrated Cavity Brillouin Optomechanics on a Chip}
\author{Yuan-Hao~Yang}
\thanks{These two authors contributed equally to this work.}
\affiliation{Laboratory of Quantum Information, University of Science and
Technology of China, Hefei 230026, P. R. China.}
\affiliation{Anhui Province Key Laboratory of Quantum Network, University of Science and Technology of China, Hefei 230026, China}

\author{Jia-Qi~Wang}
\thanks{These two authors contributed equally to this work.}
\affiliation{Laboratory of Quantum Information, University of Science and
Technology of China, Hefei 230026, P. R. China.}
\affiliation{Anhui Province Key Laboratory of Quantum Network, University of Science and Technology of China, Hefei 230026, China}

\author{Zheng-Xu~Zhu}
\affiliation{Laboratory of Quantum Information, University of Science and
Technology of China, Hefei 230026, P. R. China.}
\affiliation{Anhui Province Key Laboratory of Quantum Network, University of Science and Technology of China, Hefei 230026, China}

\author{Xin-Biao~Xu}
\email{xbxuphys@ustc.edu.cn}
\affiliation{Laboratory of Quantum Information, University of Science and
Technology of China, Hefei 230026, P. R. China.}
\affiliation{Anhui Province Key Laboratory of Quantum Network, University of Science and Technology of China, Hefei 230026, China}

\author{Ming~Li}
\affiliation{Laboratory of Quantum Information, University of Science and
Technology of China, Hefei 230026, P. R. China.}
\affiliation{Anhui Province Key Laboratory of Quantum Network, University of Science and Technology of China, Hefei 230026, China}

\author{Juanjuan~Lu}
\affiliation{School of Information Science and Technology, ShanghaiTech University, 201210 Shanghai, China}

\author{Guang-Can~Guo}
\affiliation{Laboratory of Quantum Information, University of Science and
Technology of China, Hefei 230026, P. R. China.}
\affiliation{Anhui Province Key Laboratory of Quantum Network, University of Science and Technology of China, Hefei 230026, China}
\affiliation{Hefei National Laboratory, Hefei 230088, China}

\author{Luyan~Sun}
\email{luyansun@tsinghua.edu.cn}
\affiliation{Center for Quantum Information, Institute for Interdisciplinary Information
Sciences, Tsinghua University, Beijing 100084, China}
\affiliation{Hefei National Laboratory, Hefei 230088, China}

\author{Chang-Ling~Zou}
\email{clzou321@ustc.edu.cn}
\affiliation{Laboratory of Quantum Information, University of Science and
Technology of China, Hefei 230026, P. R. China.}
\affiliation{Anhui Province Key Laboratory of Quantum Network, University of Science and Technology of China, Hefei 230026, China}
\affiliation{Hefei National Laboratory, Hefei 230088, China}

\date{\today}

\begin{abstract}
Cavity optomechanical systems enable coherent photon-phonon interactions essential for quantum technologies, yet high-performance devices have been limited to suspended structures. Here, we overcome this limitation by demonstrating cavity Brillouin optomechanics in a suspension-free racetrack microring resonator on a lithium-niobate-on-sapphire chip, a platform that merits high stability and scalability. We demonstrate coherent coupling between telecom-band optical modes and a 9.6-GHz phonon mode, achieving a maximum cooperativity of $0.41$ and a phonon quality-factor-frequency product of $10^{13}\,\mathrm{Hz}$. The momentum-matching condition inherent to traveling-wave Brillouin interactions establishes a one-to-one mapping between optical wavelength and phonon frequency, enabling multi-channel parallel operations across nearly $300\,\mathrm{MHz}$ in phonon frequency and $40\,\mathrm{nm}$ in optical wavelength. Our suspension-free architecture provides a coherent photon-phonon interface compatible with wafer-scale integration, opening pathways toward hybrid quantum circuits that unite photonic, phononic, and superconducting components on a single chip.
\end{abstract}

\maketitle

\emph{Introduction.-} Optomechanics, which explores the coherent interaction between light and mechanical motion, has emerged as a powerful platform for classical and quantum technologies~\cite{aspelmeyer_cavity_2014,barzanjeh_optomechanics_2022}. It enables fundamental studies of macroscopic quantum mechanical effects~\cite{stannigel_optomechanical_2012,marinkovic_optomechanical_2018,wallucks_quantum_2020}, precision metrology~\cite{braginsky_quantum_1980,hertzberg_back-action-evading_2010,ockeloen-korppi_quantum_2016,aasi_enhanced_2013,belenchia_testing_2016}, and microwave-to-optical frequency conversion~\cite{han_microwave-optical_2021,han_cavity_2020,mirhosseini_superconducting_2020,brubaker_optomechanical_2022,chen_optomechanical_2023,zhao_quantum-enabled_2025}. Among various optomechanical architectures, chip-integrated implementations offer dramatically enhanced coupling strengths and potential scalability through lithographic fabrication, making them particularly attractive for hybrid quantum devices~\cite{safavi2019,Clerk2020,Barzanjeh2022}. However, the full potential of phonon in optomechanics experiments remains underutilized, despite their equal role with photons as information carriers. While optical modes benefit from tight confinement in waveguides or resonators through refractive index contrast, mechanical modes typically require suspended structures to mitigate phononic dissipation (anchor loss) into the substrate~\cite{chan_laser_2011,fan_integrated_2016,otterstrom_silicon_2018,chen_optomechanical_2023,yu-chip-2025}, which compromises mechanical stability and thermal management, complicates fabrication, and critically precludes integration into scalable circuits.

Traveling-wave phononic confinement in suspension-free waveguides provides an alternative pathway by exploiting high phononic refractive index for phonon confinement, analogous to optical total internal reflection~\cite{Xu2022}. This approach aligns with Brillouin integrated photonics~\cite{eggleton_brillouin_2019,merklein_100_2022}, where stimulated scattering between traveling photon and phonon modes has enabled narrow-linewidth lasers~\cite{morrison_compact_2017,otterstrom_silicon_2018,gundavarapu_sub-hertz_2019}, microwave photonics~\cite{li_microwave_2013,marpaung_low-power_2015,ye_integrated_2025}, and non-reciprocal photonic devices~\cite{dong2015,kittlaus2018non}. By unifying these frameworks~\cite{van2016}, Brillouin optomechanics leverages momentum-matched traveling-wave interactions to achieve coherent coupling in robust, substrate-supported geometries~\cite{morrison_compact_2017,gundavarapu_sub-hertz_2019,Botter2022,Klaver2024,Neijts2024,ye_integrated_2025,Ye2025,rodrigues_cross-polarized_2025}. Nevertheless, prior implementations have been constrained by either weak phononic confinement or intrinsic dissipation in surrounding amorphous materials, hindering the high cooperative phonon-photon coupling essential for advanced applications.

In this Letter, we demonstrate cavity Brillouin optomechanics (BOM) on a lithium-niobate-on-sapphire (LNOS) platform, leveraging the traveling-wave interaction between telecom-band optical modes and a 9.6-GHz phonon mode~\cite{yang_stimulated_2023,yang2025multi}. By designing a racetrack microring resonator with its free spectral range (FSR) matched to the phonon frequency, we achieve triply-resonant enhancement, leading to a maximum cooperativity of $0.41$. Exploiting the momentum-matching condition inherent in BOM interaction~\cite{zhang_optomechanical_2017,yang_proposal_2024}, we demonstrate wideband tuning of the BOM response across a $40$\,nm optical bandwidth, corresponding to a $\sim300\,\mathrm{MHz}$ phonon frequency range. Noise power spectral density characterization reveals phonon modes with a quality (Q) factor of $1000$, yielding a high frequency-quality-factor $fQ$ product of $1.0\times 10^{13}\,\text{Hz}$. Our work establishes LNOS as a compact, robust, and high-performance platform for cavity BOM. Moreover, the inherent piezoelectricity of lithium niobate (LN) provides a pathway for hybrid electro-phononic control~\cite{Shao2022}  and coherent coupling with superconducting qubit~\cite{Hann2019,Yang2024,Xu2025QAD}, opening significant avenues for future advancement~\cite{Xu2022,xu2025magnetic}.

\begin{figure}[t]
\begin{centering}
\includegraphics[width=1\linewidth]{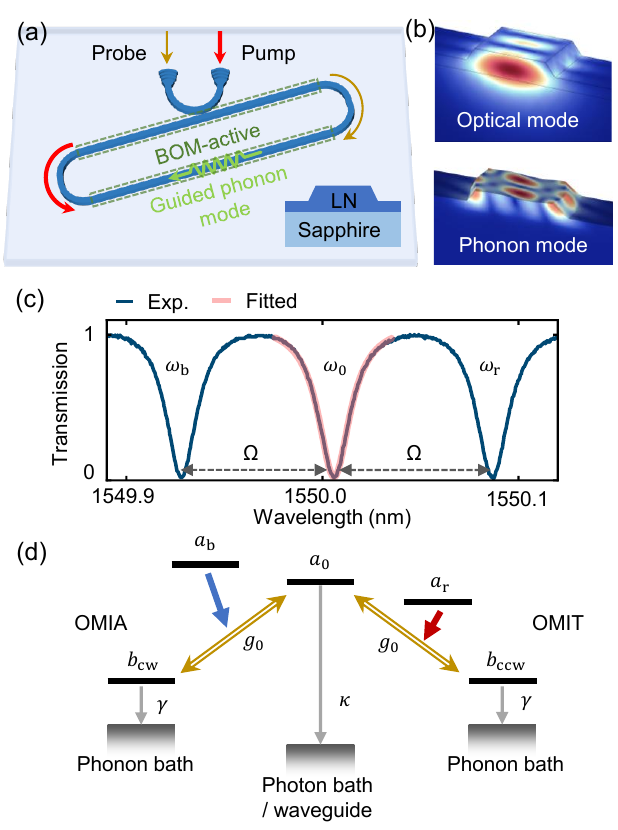}
\par\end{centering}
\caption{(a) Schematic of the racetrack microring resonator for cavity Brillouin optomechanics (BOM). Red, orange, and green arrows denote the optical pump ($a_{\text{r}}$), probe ($a_0$), and phonon ($b_{\text{cw}}$) mode, respectively. Inset: ridge waveguide cross-section. Green dashed boxes mark the straight sections deminating BOM coupling. (b) Mode profiles of traveling-wave optical (electric field) and phonon (displacement field)  modes. (c) Transmission spectrum of three participated optical modes. The red line shows the fitting result, with intrinsic and external quality factors are $2.3\times10^5$ and $2.8\times10^5$, respectively. (d) Energy level diagram for BOM interaction between optical signal ($a_0$), clockwise ($b_\mathrm{cw}$) and counter-clockwise ($b_\mathrm{ccw}$) phonon modes, under the resonantly-enhanced ($a_{\text{r}}$ and $a_{\text{b}}$) optical pump fields. OMIA/OMIT: optomechanically induced amplification/transparency; $\kappa$, $\gamma$: optical and phonon decay rates; $g_0$: vacuum BOM coupling strength.}
\label{Fig1}
\end{figure}

\emph{Device and Principle.-} Figure~\ref{Fig1}(a) illustrates our cavity BOM device, comprising a racetrack microring resonator coupled to a bus waveguide, fabricated on an X-cut LNOS chip (see~\cite{sm} for fabrication details). A cross-section of the ridge waveguide with a width of $1.2\,\mathrm{\mu m}$  and a thickness of 400\,nm thick ridge waveguide, as depicted by the inset of Fig.~\ref{Fig1}(a), featuring a carefully chosen geometry to simultaneously confine both optical and phonon modes through refractive index contrast of the LNOS platform. Figure~\ref{Fig1}(b) shows the simulated mode profiles of the fundamental transverse-electric (TE) optical mode and phonon mode. The backward Brillouin interaction arises mainly from photo-elastic and moving boundary effects~\cite{yang_proposal_2024}, enabling coherent interaction between two optical modes (pump and probe) and a phonon mode, as described by the Hamiltonian as~\cite{sipe_hamiltonian_2016,yang_proposal_2024}
\begin{align}H_{\text{I}}=g_0\left(a_{0}^{\dagger}a_{\text{r}}b_{\text{cw}}+a_{0}^{\dagger}a_{\text{b}}b_{\text{ccw}}^{\dagger}\right)+\text{h.c.},\label{eq:1}
\end{align}
where $a_{\text{r},0,\text{b}}$ and $b_{\text{cw,ccw}}$ are the bosonic operators of optical and phonon modes (with cw and ccw denoting the clockwise and counter-clockwise propagating directions), respectively, and $g_0$ is the vacuum BOM coupling strength. To maximize $g_0$, the straight sections of the racetrack resonator are oriented at an angle of 15$^{\circ}$ relative to the +Z crystal axis of the LN~\cite{yang_proposal_2024}, considering the optical and elastic anisotropy of both single-crystal LN and sapphire. Momentum-matching conditions of the backward BOM~\cite{eggleton_brillouin_2019} at telecom wavelengths yield phonon frequencies of $\sim9\,\mathrm{GHz}$ on the LNOS platform~\cite{yang_stimulated_2023,yang_proposal_2024,yang2025multi}. The perimeter of the racetrack resonator is optimized to 13.43\,mm, resulting in an FSR of $9.79\,\mathrm{GHz}$ that matches the phonon frequencies. Figure ~\ref{Fig1}(c)  shows a typical optical transmission spectrum~\cite{sm}. For a selected probe mode $a_0$, the pump can drive on-resonance with either the lower-frequency mode ($\omega_{\text{r}}$, red-detuned) or higher-frequency mode ($\omega_{\text{b}}$, blue-detuned), leading to resonantly enhanced BOM that achieves optomechanically induced transparency (OMIT) or  amplification (OMIA), respectively, as illustrated in Fig.~\ref{Fig1}(d).

\begin{figure*}[t]
\begin{centering}
\includegraphics[width=1\linewidth]{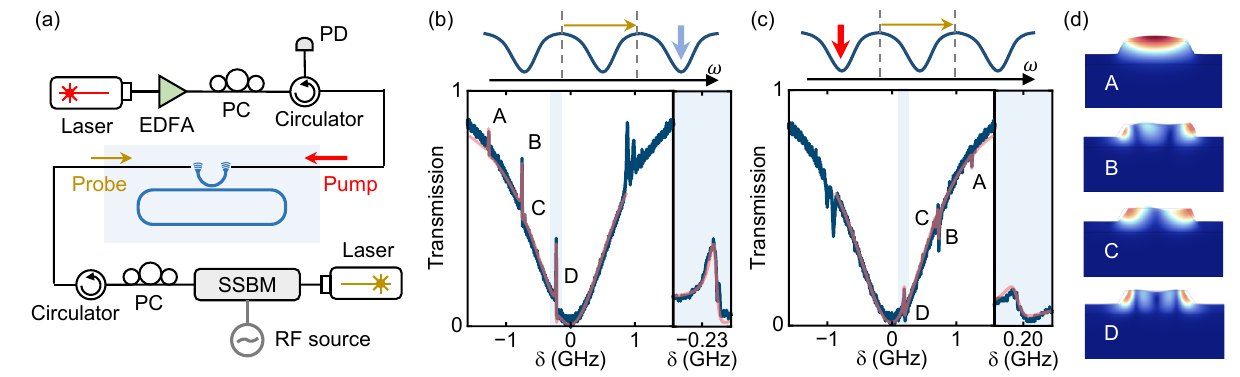}
\par\end{centering}
\caption{(a) Experimental setup for BOM characterization. EDFA: erbium-doped fiber amplifier; PC: polarization controller; PD: photo-detector; RF source: radio-frequency source; SSBM: single-sideband modulator. (b) and (c) Schematics and measured spectra of OMIA and OMIT, respectively, with an on-chip pump power of $121\,\mathrm{mW}$. Red lines denote the fits; red, blue, and orange arrows represent OMIA pump, OMIT pump, and probe light, respectively; gray dashed lines indicate the probe sweep range. (d) Mode profiles of phonon modes $A$, $B$, $C$ and $D$, corresponding to labeled peaks in (b) and (c).}
\label{Fig2}
\end{figure*}

\emph{OMIA and OMIT.-} The BOM interaction is experimentally characterized by the setup shown in Fig~\ref{Fig2}(a). The pump laser is amplified by an erbium-doped fiber amplifier (EDFA) and injected into the device via a grating coupler~\cite{chen2025fiber}, tuned to on-resonance with a selected optical mode, while a weak probe laser is input to the device from counter-propagating direction. Optical circulators are used to route the transmitted probe signal to a photo-detector (PD), while simultaneously protecting the EDFA from the strong backscattered pump light. Employing a single-sideband modulator (SSBM), high-resolution probe transmission spectra across probe mode ($a_0$) are obtained through sweeping the radio-frequency (RF) signal applied to the SSBM. The probe power transmission is given by~\cite{sm}
\begin{align}
    T = \left|1 +\frac{2\kappa_{\text{ex}}}{-\text{i}\delta-\kappa-\frac{|G|^2}{i\Delta\pm\gamma}}\right|^2,\label{eq:2}
\end{align}
where $+/-$ denotes the pump on red/blue detuned modes, corresponding to the OMIT and OMIA, respectively, $\delta=\omega_0-\omega_{\text{s}}$ is the probe the detuning, $\Delta=\Omega-\left|\omega_\text{s}-\omega_{\text{p}}\right|$ denotes the frequency mismatch for the triply-resonant BOM, with $\omega_0$ and $\Omega$ being optical and phonon mode frequencies [Fig.~\ref{Fig1}(c)], $\omega_\text{s}$ and $\omega_{\text{p}}$ being probe and pump frequencies, and $G=\sqrt{n_{\mathrm{p}}} g_0$ is the pump stimulated coupling strength under a intracavity pump photon number $n_{\mathrm{p}}$.  The cooperativity $C = |G|^2/\kappa\gamma$ is a figure of merit that quantifies the coherent interaction, with $\kappa$ and $\gamma$ denoting the amplitude decay rates of the optical and phonon modes, respectively. When the probe frequency meets BOM resonant condition ($\Delta=0$), the coherent interconversion between photon and phonon modifies the decay rate of the photons to $(1\pm C)\kappa$, which in turn alters the corresponding extinction ratio of the probe transmission.

Figures~\ref{Fig2}(b) and (c) presents the measured OMIA and OMIT spectra. For OMIA [Fig.~\ref{Fig2}(b)], the pump resonant with $a_{\text{b}}$ (blue arrow), while the probe sweeps across $a_{0}$ (yellow arrow). Compared with the spectrum without pump [Fig.~\ref{Fig1}(b)], a series of sharp peaks arises on the resonance dip can be categorized into two groups: three on the right are from stimulated Brillouin scattering in the single-mode optical fiber, while the four on the left, labeled as $A$, $B$, $C$, and $D$, are OMIA signals generated by different phonon modes in the waveguide [Fig.~\ref{Fig2}(d)]. A detailed view of the OMIA signal of mode $D$ (shaded region) reveals a narrow amplification window with a full width at half maximum (FWHM) of $\sim9$ MHz, highlighting the system's potential as a narrow-band, non-reciprocal optical amplifier~
\cite{Shen2018}. Similarly, for OMIT [Fig.~\ref{Fig2}(c)], the pump resonant with $a_{\text{r}}$ (red arrow), generating multiple sharp dips on the spectrum. It is worth noting that both OMIT and OMIA can induce either dips or peaks in the transmission spectrum, depending on whether the probe mode operates in the over-coupling or under-coupling regime; however, only OMIA can produce amplified transmission ($T>1$) when $C$ is sufficiently large.

\begin{figure}[b]
\begin{centering}
\includegraphics[width=1\linewidth]{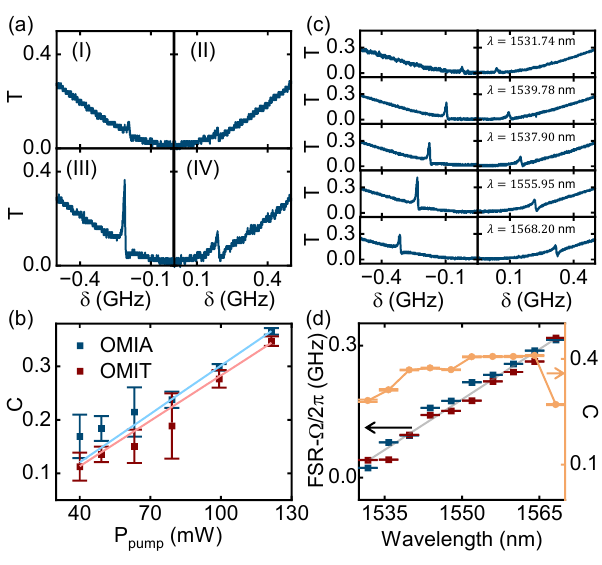}
\par\end{centering}
\caption{(a) OMIA and OMIT spectra with different pump powers. On-chip $P_{\text{pump}} = 40.01\,\text{mW}$ for (I) and (II), and $121.41\,\text{mW}$ for (III) and (IV). (b) Dependence between cooperativity $C$ and $P_{\text{pump}}$ extracted from OMIA and OMIT spectra, respectively. The dots represent the fitted $C$, and the error bars represent the corresponding fitting errors. The solid lines represent the linear fits. (c) OMIA and OMIT spectra of modes with different optical wavelength $\lambda$. (d) Frequency mismatch between FSR and phonon mode ($\Omega$) and corresponding $C$ vary with optical wavelength. Blue and red dots: the fitted results of OMIA and OMIT, respectively. Solid line: a linear fit. Orange dots: fitted C factor.}
\label{Fig3}
\end{figure}

\begin{figure*}[t]
\begin{centering}
\includegraphics[width=1\linewidth]{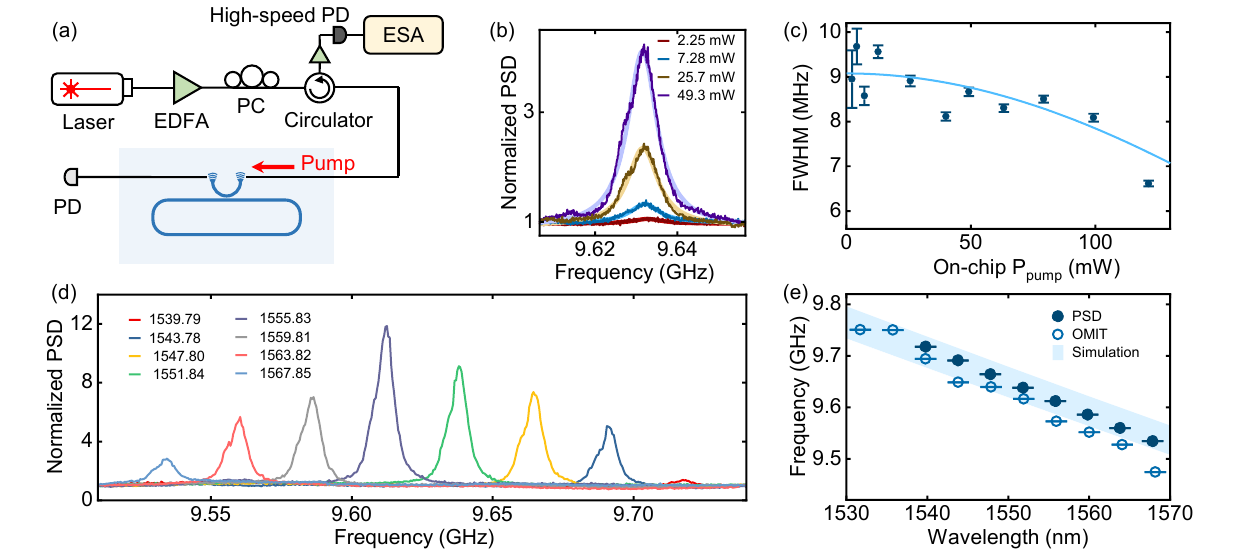}
\par\end{centering}
\caption{(a) Experimental setup for noise power spectrum density (PSD) characterization. ESA: electrical spectrum analyzer. (b) Normalized PSD of the phonon mode with varied $P_{\text{pump}}$. Solid lines with dark and light colors represent experimental and fitting results, respectively. (c) Full-width-at-half-maximum (FWHM) varies with $P_{\text{pump}}$. The blue dots and solid line represent the experimental and fitting results, respectively. (d) Normalized PSD with different optical wavelength. (e) Dependence of the phonon mode frequency on optical wavelength, extracted from noise PSD spectra (solid blue dots) and OMIT spectra (hollow blue dots), respectively. The shaded region corresponds to the simulated results, for which we calculate structures with a waveguide width of $1.2\pm0.1\,\mu \text{m}$ to factor in experimental fabrication tolerances.}
\label{Fig4}
\end{figure*}

\textit{Wavelength-frequency mapping.-} We select the phonon mode $D$, which exhibits the largest $C$ and the smallest $\Delta$, for further investigation and the results are summarized in Fig.~\ref{Fig3}. By increasing the pump power  $P_\mathrm{pump}$, i.e., larger $n_\mathrm{p}\propto P_\mathrm{pump}$, the spectral features of OMIA and OMIT become more pronounced [Fig.~\ref{Fig3}(a)] as expected. By fitting the spectral lineshapes at various pump powers using Eq.~(\ref{eq:2}), $C$ is extracted and plotted in Fig.~\ref{Fig3}(b). The observed linearly increased $C$ with the on-chip pump power agrees with the prediction $C=n_\mathrm{p}g_0^2/\kappa\gamma\propto P_{\mathrm{pump}}$. With experimentally calibrated parameters, we derived $g_0/2\pi = 2.69\pm0.29\,\mathrm{kHz}$ for OMIA and $g_0/2\pi = 2.61\pm0.29\,\mathrm{kHz}$ for OMIT.

A key advantage of BOM compared with conventional optomechanics is the traveling-wave nature of the interaction, which allows BOM interaction between a group of optical and phonon modes~\cite{Grudinin2009,Lee2012}. Consequently, the momentum-matching condition leads to a direct correspondence between the optical wavelength ($\lambda$) and the phonon frequency ($\Omega$) as $\frac{\Delta \Omega/2\pi}{\Delta \lambda} = -\frac{2n_{\text{g}}}{\lambda^2v_{\text{g}}}$, where $n_{\text{g}}$ and $v_{\text{g}}$ are the optical group refractive index and phononic group velocity, respectively~\cite{sm}. We experimentally verify this relation by performing OMIA/OMIT measurements across a broad optical wavelength range. As shown in Figs.~\ref{Fig3}(c), the frequency mismatch between the phonon and optical modes increases at longer wavelengths. The extracted frequency mismatch ($\mathrm{FSR}-\Omega/2\pi$), as plotted in Fig.~\ref{Fig3}(d), exhibits a linear dependence on $\lambda$ with a fitted slope of $7.60\pm0.29\,\mathrm{MHz/nm}$. Due to the limited bandwidths of the grating couplers and EDFA, the extracted $C$ also varies with $\lambda$ [Fig.~\ref{Fig3}(d)], with a maximum of $C=0.410\pm0.001$ achieved.

\emph{Noise power spectral density.-} To independently characterize the phonon mode and confirm the multi-channel capability, we measure the noise power spectral density (PSD)~\cite{chan_laser_2011,cohen_phonon_2015}. As illustrated in Fig.~\ref{Fig4}(a), the setup employs only a pump input to stimulate BOM interaction with thermally excited phonons. The generated blue-shifted and red-shifted photons, corresponding to anti-Stokes and Stokes processes respectively, along with the reflected pump, are amplified by a low-noise EDFA and detected through heterodyning detection on a high-speed PD. The pump serves as a local oscillator (LO), and the resulting photocurrent beat note is analyzed using an electrical spectrum analyzer (ESA). The normalized noise PSD of the photocurrent is given by~\cite{sm}
\begin{equation}
    S_{II}(\omega) = 1 + \frac{8\kappa_\text{ex}|G|^2}{\kappa^2}\left[S_{\text{b}}(\omega)+S_{\text{b}}(-\omega)+S_{\text{r}}(\omega)+S_{\text{r}}(-\omega)\right],\label{eq:4} \nonumber
    \end{equation}
with contributions $S_{\text{b}}(\omega) = \frac{n_\text{th}+1}{1-C}\frac{(1-C)\gamma}{(\Omega-\omega)^2+(1-C)^2\gamma^2}$ from OMIA and $S_{\text{r}}(\omega) = \frac{n_\text{th}}{1+C}\frac{(1+C)\gamma}{(\Omega-\omega)^2+(1+C)^2\gamma^2}$ from OMIT, where $n_{\text{th}}$ is the thermal phonon occupancy and $\kappa_\mathrm{ex}$ is the external coupling rate of the optical modes. Figure~\ref{Fig4}(b) shows typical noise PSD at different pump powers. By fitting these spectra [Fig.~\ref{Fig4}(c)], we extract an intrinsic phonon amplitude decay rate of $\gamma/2\pi = 4.54\pm0.09\,\text{MHz}$, corresponding to a quality factor of $\Omega/2\gamma\approx 1000$. By selecting different optical modes for the pump, we obtain the corresponding PSDs shown in Fig.~\ref{Fig4}(d). The peak amplitude of the PSD decreases as the wavelength deviates from the center ($\sim 1550\,\mathrm{nm}$), attributed to the limited bandwidth of the grating couplers as the $C$ diminishes. The dependence of the phonon frequency on $\lambda$ [solid dots in Fig.~\ref{Fig4}(e)] shows excellent agreement with both the OMIT measurement (hollow circles) and numerical simulations (shaded region), showing a linear dependence spanning a wavelength range from 1530~nm to 1570~nm with corresponding phonon frequencies from 9.5~GHz to 9.8~GHz. These results confirm flexible mode selection for high-cooperativity BOM, enabling multi-channel parallel operation over a wide spectral range.

\emph{Conclusion.-} We have demonstrated a suspension-free cavity Brillouin optomechanical system on the LNOS platform. A vacuum photon-phonon coupling strength of $g_0/2\pi=2.69\,\text{kHz}$ and a high cooperativity of $C=0.41$ are achieved, benefit from the high performance phonon modes with a high $fQ$-product of $10^{13}\,\mathrm{Hz}$ in our hybrid photon-phonon platform. It is anticipated that the performances can be further improved by operating the device at cryogenic temperatures~\cite{xu_high-frequency_2022}. Our experiment demonstrates a one-to-one mapping between optical wavelength and phonon frequency, providing an intrinsic multi-channel capabilities for parallel operations across nearly $300\,\mathrm{MHz}$ phonon frequency bandwidth. This robust, high-performance, and multi-channel system represents a significant step for integrated phononic-photonic devices. Furthermore, the inherent piezoelectricity of LN provides a direct electrical interface for coherent control of traveling-wave phonons, making this platform highly promising for hybrid quantum systems~\cite{Xu2022,Xu2025QAD}, quantum phononics~\cite{manenti_circuit_2017,Hann2019,Yang2024}, and quantum frequency conversion~\cite{han_microwave-optical_2021}.

\smallskip{}
\begin{acknowledgments}
This work was funded by the National Natural Science Foundation of China (Grant Nos.~92265210, 92265108, 123B2068, 92165209, 92365301, 12474498, 11925404, 12374361, and 12293053), the Innovation Program for Quantum Science and Technology (Grant Nos.~2021ZD0300200 and 2024ZD0301500). We also acknowledge the support from the Fundamental Research Funds for the Central Universities and USTC Research Funds of the Double First-Class Initiative. The numerical calculations in this paper were performed on the supercomputing system in the Supercomputing Center of University of Science and Technology of China. This work was partially carried out at the USTC Center for Micro and Nanoscale Research and Fabrication.
\end{acknowledgments}


\begin{thebibliography}{57}%
\makeatletter
\providecommand \@ifxundefined [1]{%
 \@ifx{#1\undefined}
}%
\providecommand \@ifnum [1]{%
 \ifnum #1\expandafter \@firstoftwo
 \else \expandafter \@secondoftwo
 \fi
}%
\providecommand \@ifx [1]{%
 \ifx #1\expandafter \@firstoftwo
 \else \expandafter \@secondoftwo
 \fi
}%
\providecommand \natexlab [1]{#1}%
\providecommand \enquote  [1]{``#1''}%
\providecommand \bibnamefont  [1]{#1}%
\providecommand \bibfnamefont [1]{#1}%
\providecommand \citenamefont [1]{#1}%
\providecommand \href@noop [0]{\@secondoftwo}%
\providecommand \href [0]{\begingroup \@sanitize@url \@href}%
\providecommand \@href[1]{\@@startlink{#1}\@@href}%
\providecommand \@@href[1]{\endgroup#1\@@endlink}%
\providecommand \@sanitize@url [0]{\catcode `\\12\catcode `\$12\catcode
  `\&12\catcode `\#12\catcode `\^12\catcode `\_12\catcode `\%12\relax}%
\providecommand \@@startlink[1]{}%
\providecommand \@@endlink[0]{}%
\providecommand \url  [0]{\begingroup\@sanitize@url \@url }%
\providecommand \@url [1]{\endgroup\@href {#1}{\urlprefix }}%
\providecommand \urlprefix  [0]{URL }%
\providecommand \Eprint [0]{\href }%
\providecommand \doibase [0]{http://dx.doi.org/}%
\providecommand \selectlanguage [0]{\@gobble}%
\providecommand \bibinfo  [0]{\@secondoftwo}%
\providecommand \bibfield  [0]{\@secondoftwo}%
\providecommand \translation [1]{[#1]}%
\providecommand \BibitemOpen [0]{}%
\providecommand \bibitemStop [0]{}%
\providecommand \bibitemNoStop [0]{.\EOS\space}%
\providecommand \EOS [0]{\spacefactor3000\relax}%
\providecommand \BibitemShut  [1]{\csname bibitem#1\endcsname}%
\let\auto@bib@innerbib\@empty
\bibitem [{\citenamefont {Aspelmeyer}\ \emph {,~et~al.}(2014)\citenamefont
  {Aspelmeyer}, \citenamefont {Kippenberg},\ and\ \citenamefont
  {Marquardt}}]{aspelmeyer_cavity_2014}%
  \BibitemOpen
  \bibfield  {author} {\bibinfo {author} {\bibfnamefont {M.}~\bibnamefont
  {Aspelmeyer}}, \bibinfo {author} {\bibfnamefont {T.~J.}\ \bibnamefont
  {Kippenberg}}, \ and\ \bibinfo {author} {\bibfnamefont {F.}~\bibnamefont
  {Marquardt}},\ }\bibfield  {title} {\bibinfo {title} {Cavity optomechanics},\
  }\href {\doibase10.1103/RevModPhys.86.1391} {\bibfield  {journal} {\bibinfo
  {journal} {Rev. Mod. Phys.}\ }\textbf {\bibinfo {volume} {86}},\ \bibinfo
  {pages} {1391} (\bibinfo {year} {2014})}\BibitemShut {NoStop}%
\bibitem [{\citenamefont {Barzanjeh}\ \emph
  {,~et~al.}(2022{\natexlab{a}})\citenamefont {Barzanjeh}, \citenamefont
  {Xuereb}, \citenamefont {Groblacher}, \citenamefont {Paternostro},
  \citenamefont {Regal},\ and\ \citenamefont
  {Weig}}]{barzanjeh_optomechanics_2022}%
  \BibitemOpen
  \bibfield  {author} {\bibinfo {author} {\bibfnamefont {S.}~\bibnamefont
  {Barzanjeh}}, \bibinfo {author} {\bibfnamefont {A.}~\bibnamefont {Xuereb}},
  \bibinfo {author} {\bibfnamefont {S.}~\bibnamefont {Groblacher}}, \bibinfo
  {author} {\bibfnamefont {M.}~\bibnamefont {Paternostro}}, \bibinfo {author}
  {\bibfnamefont {C.~A.}\ \bibnamefont {Regal}}, \ and\ \bibinfo {author}
  {\bibfnamefont {E.~M.}\ \bibnamefont {Weig}},\ }\bibfield  {title} {\bibinfo
  {title} {Optomechanics for quantum technologies},\ }\href
  {\doibase10.1038/s41567-021-01402-0} {\bibfield  {journal} {\bibinfo
  {journal} {Nat. Phys.}\ }\textbf {\bibinfo {volume} {18}},\ \bibinfo {pages}
  {15} (\bibinfo {year} {2022}{\natexlab{a}})}\BibitemShut {NoStop}%
\bibitem [{\citenamefont {Stannigel}\ \emph {,~et~al.}(2012)\citenamefont
  {Stannigel}, \citenamefont {Komar}, \citenamefont {Habraken}, \citenamefont
  {Bennett}, \citenamefont {Lukin}, \citenamefont {Zoller},\ and\ \citenamefont
  {Rabl}}]{stannigel_optomechanical_2012}%
  \BibitemOpen
  \bibfield  {author} {\bibinfo {author} {\bibfnamefont {K.}~\bibnamefont
  {Stannigel}}, \bibinfo {author} {\bibfnamefont {P.}~\bibnamefont {Komar}},
  \bibinfo {author} {\bibfnamefont {S.~J.~M.}\ \bibnamefont {Habraken}},
  \bibinfo {author} {\bibfnamefont {S.~D.}\ \bibnamefont {Bennett}}, \bibinfo
  {author} {\bibfnamefont {M.~D.}\ \bibnamefont {Lukin}}, \bibinfo {author}
  {\bibfnamefont {P.}~\bibnamefont {Zoller}}, \ and\ \bibinfo {author}
  {\bibfnamefont {P.}~\bibnamefont {Rabl}},\ }\bibfield  {title} {\bibinfo
  {title} {Optomechanical {Quantum} {Information} {Processing} with {Photons}
  and {Phonons}},\ }\href {\doibase10.1103/PhysRevLett.109.013603} {\bibfield
  {journal} {\bibinfo  {journal} {Phys. Rev. Lett.}\ }\textbf {\bibinfo
  {volume} {109}},\ \bibinfo {pages} {013603} (\bibinfo {year}
  {2012})}\BibitemShut {NoStop}%
\bibitem [{\citenamefont {Marinkovic}\ \emph {,~et~al.}(2018)\citenamefont
  {Marinkovic}, \citenamefont {Wallucks}, \citenamefont {Riedinger},
  \citenamefont {Hong}, \citenamefont {Aspelmeyer},\ and\ \citenamefont
  {Groblacher}}]{marinkovic_optomechanical_2018}%
  \BibitemOpen
  \bibfield  {author} {\bibinfo {author} {\bibfnamefont {I.}~\bibnamefont
  {Marinkovic}}, \bibinfo {author} {\bibfnamefont {A.}~\bibnamefont
  {Wallucks}}, \bibinfo {author} {\bibfnamefont {R.}~\bibnamefont {Riedinger}},
  \bibinfo {author} {\bibfnamefont {S.}~\bibnamefont {Hong}}, \bibinfo {author}
  {\bibfnamefont {M.}~\bibnamefont {Aspelmeyer}}, \ and\ \bibinfo {author}
  {\bibfnamefont {S.}~\bibnamefont {Groblacher}},\ }\bibfield  {title}
  {\bibinfo {title} {Optomechanical {Bell} {Test}},\ }\href
  {\doibase10.1103/PhysRevLett.121.220404} {\bibfield  {journal} {\bibinfo
  {journal} {Phys. Rev. Lett.}\ }\textbf {\bibinfo {volume} {121}},\ \bibinfo
  {pages} {220404} (\bibinfo {year} {2018})}\BibitemShut {NoStop}%
\bibitem [{\citenamefont {Wallucks}\ \emph {,~et~al.}(2020)\citenamefont
  {Wallucks}, \citenamefont {Marinkovic}, \citenamefont {Hensen}, \citenamefont
  {Stockill},\ and\ \citenamefont {Groblacher}}]{wallucks_quantum_2020}%
  \BibitemOpen
  \bibfield  {author} {\bibinfo {author} {\bibfnamefont {A.}~\bibnamefont
  {Wallucks}}, \bibinfo {author} {\bibfnamefont {I.}~\bibnamefont
  {Marinkovic}}, \bibinfo {author} {\bibfnamefont {B.}~\bibnamefont {Hensen}},
  \bibinfo {author} {\bibfnamefont {R.}~\bibnamefont {Stockill}}, \ and\
  \bibinfo {author} {\bibfnamefont {S.}~\bibnamefont {Groblacher}},\ }\bibfield
   {title} {\bibinfo {title} {A quantum memory at telecom wavelengths},\ }\href
  {\doibase10.1038/s41567-020-0891-z} {\bibfield  {journal} {\bibinfo
  {journal} {Nat. Phys.}\ }\textbf {\bibinfo {volume} {16}},\ \bibinfo {pages}
  {772} (\bibinfo {year} {2020})}\BibitemShut {NoStop}%
\bibitem [{\citenamefont {Braginsky}\ \emph {,~et~al.}(1980)\citenamefont
  {Braginsky}, \citenamefont {Vorontsov},\ and\ \citenamefont
  {Thorne}}]{braginsky_quantum_1980}%
  \BibitemOpen
  \bibfield  {author} {\bibinfo {author} {\bibfnamefont {V.~B.}\ \bibnamefont
  {Braginsky}}, \bibinfo {author} {\bibfnamefont {Y.~I.}\ \bibnamefont
  {Vorontsov}}, \ and\ \bibinfo {author} {\bibfnamefont {K.~S.}\ \bibnamefont
  {Thorne}},\ }\bibfield  {title} {\bibinfo {title} {Quantum {Nondemolition}
  {Measurements}},\ }\href {\doibase10.1126/science.209.4456.547} {\bibfield
  {journal} {\bibinfo  {journal} {Science}\ }\textbf {\bibinfo {volume}
  {209}},\ \bibinfo {pages} {547} (\bibinfo {year} {1980})}\BibitemShut
  {NoStop}%
\bibitem [{\citenamefont {Hertzberg}\ \emph {,~et~al.}(2010)\citenamefont
  {Hertzberg}, \citenamefont {Rocheleau}, \citenamefont {Ndukum}, \citenamefont
  {Savva}, \citenamefont {Clerk},\ and\ \citenamefont
  {Schwab}}]{hertzberg_back-action-evading_2010}%
  \BibitemOpen
  \bibfield  {author} {\bibinfo {author} {\bibfnamefont {J.~B.}\ \bibnamefont
  {Hertzberg}}, \bibinfo {author} {\bibfnamefont {T.}~\bibnamefont
  {Rocheleau}}, \bibinfo {author} {\bibfnamefont {T.}~\bibnamefont {Ndukum}},
  \bibinfo {author} {\bibfnamefont {M.}~\bibnamefont {Savva}}, \bibinfo
  {author} {\bibfnamefont {A.~A.}\ \bibnamefont {Clerk}}, \ and\ \bibinfo
  {author} {\bibfnamefont {K.~C.}\ \bibnamefont {Schwab}},\ }\bibfield  {title}
  {\bibinfo {title} {Back-action-evading measurements of nanomechanical
  motion},\ }\href {\doibase10.1038/nphys1479} {\bibfield  {journal} {\bibinfo
  {journal} {Nat. Phys.}\ }\textbf {\bibinfo {volume} {6}},\ \bibinfo {pages}
  {213} (\bibinfo {year} {2010})}\BibitemShut {NoStop}%
\bibitem [{\citenamefont {Ockeloen-Korppi}\ \emph {,~et~al.}(2016)\citenamefont
  {Ockeloen-Korppi}, \citenamefont {Damskagg}, \citenamefont {Pirkkalainen},
  \citenamefont {Clerk}, \citenamefont {Woolley},\ and\ \citenamefont
  {Sillanpaa}}]{ockeloen-korppi_quantum_2016}%
  \BibitemOpen
  \bibfield  {author} {\bibinfo {author} {\bibfnamefont {C.}~\bibnamefont
  {Ockeloen-Korppi}}, \bibinfo {author} {\bibfnamefont {E.}~\bibnamefont
  {Damskagg}}, \bibinfo {author} {\bibfnamefont {J.-M.}\ \bibnamefont
  {Pirkkalainen}}, \bibinfo {author} {\bibfnamefont {A.}~\bibnamefont {Clerk}},
  \bibinfo {author} {\bibfnamefont {M.}~\bibnamefont {Woolley}}, \ and\
  \bibinfo {author} {\bibfnamefont {M.}~\bibnamefont {Sillanpaa}},\ }\bibfield
  {title} {\bibinfo {title} {Quantum {Backaction} {Evading} {Measurement} of
  {Collective} {Mechanical} {Modes}},\ }\href
  {\doibase10.1103/PhysRevLett.117.140401} {\bibfield  {journal} {\bibinfo
  {journal} {Phys. Rev. Lett.}\ }\textbf {\bibinfo {volume} {117}},\ \bibinfo
  {pages} {140401} (\bibinfo {year} {2016})}\BibitemShut {NoStop}%
\bibitem [{\citenamefont {Aasi}\ \emph {,~et~al.}(2013)\citenamefont {Aasi},
  \citenamefont {Abadie}, \citenamefont {Abbott}, \citenamefont {Abbott},
  \citenamefont {Abbott}, \citenamefont {Abernathy}, \citenamefont {Adams},
  \citenamefont {Adams}, \citenamefont {Addesso}, \citenamefont {Adhikari},
  \citenamefont {Affeldt}, \citenamefont {Aguiar}, \citenamefont {Ajith},
  \citenamefont {Allen}, \citenamefont {Amador~Ceron}, \citenamefont
  {Amariutei}, \citenamefont {Anderson}, \citenamefont {Anderson},
  \citenamefont {Arai}, \citenamefont {Araya}, \citenamefont {Arceneaux},
  \citenamefont {Ast}, \citenamefont {Aston}, \citenamefont {Atkinson},
  \citenamefont {Aufmuth}, \citenamefont {Aulbert}, \citenamefont {Austin},
  \citenamefont {Aylott}, \citenamefont {Babak}, \citenamefont {Baker},
  \citenamefont {Ballmer}, \citenamefont {Bao}, \citenamefont {Barayoga},
  \citenamefont {Barker}, \citenamefont {Barr}, \citenamefont {Barsotti},
  \citenamefont {Barton}, \citenamefont {Bartos}, \citenamefont {Bassiri},
  \citenamefont {Batch}, \citenamefont {Bauchrowitz}, \citenamefont {Behnke},
  \citenamefont {Bell}, \citenamefont {Bell}, \citenamefont {Bergmann},
  \citenamefont {Berliner}, \citenamefont {Bertolini}, \citenamefont
  {Betzwieser}, \citenamefont {Beveridge}, \citenamefont {Beyersdorf},
  \citenamefont {Bhadbhade}, \citenamefont {Bilenko}, \citenamefont
  {Billingsley}, \citenamefont {Birch}, \citenamefont {Biscans}, \citenamefont
  {Black}, \citenamefont {Blackburn}, \citenamefont {Blackburn}, \citenamefont
  {Blair}, \citenamefont {Bland}, \citenamefont {Bock}, \citenamefont {Bodiya},
  \citenamefont {Bogan}, \citenamefont {Bond}, \citenamefont {Bork},
  \citenamefont {Born}, \citenamefont {Bose}, \citenamefont {Bowers},
  \citenamefont {Brady}, \citenamefont {Braginsky}, \citenamefont {Brau},
  \citenamefont {Breyer}, \citenamefont {Bridges}, \citenamefont {Brinkmann},
  \citenamefont {Britzger}, \citenamefont {Brooks}, \citenamefont {Brown},
  \citenamefont {Brown}, \citenamefont {Buckland}, \citenamefont {Brš¹ckner},
  \citenamefont {Buchler}, \citenamefont {Buonanno}, \citenamefont
  {Burguet-Castell}, \citenamefont {Byer}, \citenamefont {Cadonati},
  \citenamefont {Camp}, \citenamefont {Campsie}, \citenamefont {Cannon},
  \citenamefont {Cao}, \citenamefont {Capano}, \citenamefont {Carbone},
  \citenamefont {Caride}, \citenamefont {Castiglia}, \citenamefont {Caudill},
  \citenamefont {Cavaglia}, \citenamefont {Cepeda}, \citenamefont
  {Chalermsongsak}, \citenamefont {Chao}, \citenamefont {Charlton},
  \citenamefont {Chen}, \citenamefont {Chen}, \citenamefont {Cho},
  \citenamefont {Chow}, \citenamefont {Christensen}, \citenamefont {Chu},
  \citenamefont {Chua}, \citenamefont {Chung}, \citenamefont {Ciani},
  \citenamefont {Clara}, \citenamefont {Clark}, \citenamefont {Clark},
  \citenamefont {Constancio~Junior}, \citenamefont {Cook}, \citenamefont
  {Corbitt}, \citenamefont {Cordier}, \citenamefont {Cornish}, \citenamefont
  {Corsi}, \citenamefont {Costa}, \citenamefont {Coughlin}, \citenamefont
  {Countryman}, \citenamefont {Couvares}, \citenamefont {Coward}, \citenamefont
  {Cowart}, \citenamefont {Coyne}, \citenamefont {Craig}, \citenamefont
  {Creighton}, \citenamefont {Creighton}, \citenamefont {Cumming},
  \citenamefont {Cunningham}, \citenamefont {Dahl}, \citenamefont {Damjanic},
  \citenamefont {Danilishin}, \citenamefont {Danzmann}, \citenamefont
  {Daudert}, \citenamefont {Daveloza}, \citenamefont {Davies}, \citenamefont
  {Daw}, \citenamefont {Dayanga}, \citenamefont {Deleeuw}, \citenamefont
  {Denker}, \citenamefont {Dent}, \citenamefont {Dergachev}, \citenamefont
  {DeRosa}, \citenamefont {DeSalvo}, \citenamefont {Dhurandhar}, \citenamefont
  {Di~Palma}, \citenamefont {Dšªaz}, \citenamefont {Dietz}, \citenamefont
  {Donovan}, \citenamefont {Dooley}, \citenamefont {Doravari}, \citenamefont
  {Drasco}, \citenamefont {Drever}, \citenamefont {Driggers}, \citenamefont
  {Du}, \citenamefont {Dumas}, \citenamefont {Dwyer}, \citenamefont {Eberle},
  \citenamefont {Edwards}, \citenamefont {Effler}, \citenamefont {Ehrens},
  \citenamefont {Eikenberry}, \citenamefont {Engel}, \citenamefont {Essick},
  \citenamefont {Etzel}, \citenamefont {Evans}, \citenamefont {Evans},
  \citenamefont {Evans}, \citenamefont {Factourovich}, \citenamefont
  {Fairhurst}, \citenamefont {Fang}, \citenamefont {Farr}, \citenamefont
  {Farr}, \citenamefont {Favata}, \citenamefont {Fazi}, \citenamefont
  {Fehrmann}, \citenamefont {Feldbaum}, \citenamefont {Finn}, \citenamefont
  {Fisher}, \citenamefont {Foley}, \citenamefont {Forsi}, \citenamefont
  {Fotopoulos}, \citenamefont {Frede}, \citenamefont {Frei}, \citenamefont
  {Frei}, \citenamefont {Freise}, \citenamefont {Frey}, \citenamefont {Fricke},
  \citenamefont {Friedrich}, \citenamefont {Fritschel}, \citenamefont {Frolov},
  \citenamefont {Fujimoto}, \citenamefont {Fulda}, \citenamefont {Fyffe},
  \citenamefont {Gair}, \citenamefont {Garcia}, \citenamefont {Gehrels},
  \citenamefont {Gelencser}, \citenamefont {Gergely}, \citenamefont {Ghosh},
  \citenamefont {Giaime}, \citenamefont {Giampanis}, \citenamefont {Giardina},
  \citenamefont {Gil-Casanova}, \citenamefont {Gill}, \citenamefont {Gleason},
  \citenamefont {Goetz}, \citenamefont {Gonzš¢lez}, \citenamefont {Gordon},
  \citenamefont {Gorodetsky}, \citenamefont {Gossan}, \citenamefont {Go?ler},
  \citenamefont {Graef}, \citenamefont {Graff}, \citenamefont {Grant},
  \citenamefont {Gras}, \citenamefont {Gray}, \citenamefont {Greenhalgh},
  \citenamefont {Gretarsson}, \citenamefont {Griffo}, \citenamefont {Grote},
  \citenamefont {Grover}, \citenamefont {Grunewald}, \citenamefont {Guido},
  \citenamefont {Gustafson}, \citenamefont {Gustafson}, \citenamefont {Hammer},
  \citenamefont {Hammond}, \citenamefont {Hanks}, \citenamefont {Hanna},
  \citenamefont {Hanson}, \citenamefont {Haris}, \citenamefont {Harms},
  \citenamefont {Harry}, \citenamefont {Harry}, \citenamefont {Harstad},
  \citenamefont {Hartman}, \citenamefont {Haughian}, \citenamefont {Hayama},
  \citenamefont {Heefner}, \citenamefont {Heintze}, \citenamefont {Hendry},
  \citenamefont {Heng}, \citenamefont {Heptonstall}, \citenamefont {Heurs},
  \citenamefont {Hewitson}, \citenamefont {Hild}, \citenamefont {Hoak},
  \citenamefont {Hodge}, \citenamefont {Holt}, \citenamefont {Holtrop},
  \citenamefont {Hong}, \citenamefont {Hooper}, \citenamefont {Hough},
  \citenamefont {Howell}, \citenamefont {Huang}, \citenamefont {Huerta},
  \citenamefont {Hughey}, \citenamefont {Huttner}, \citenamefont {Huynh},
  \citenamefont {Huynh-Dinh}, \citenamefont {Ingram}, \citenamefont {Inta},
  \citenamefont {Isogai}, \citenamefont {Ivanov}, \citenamefont {Iyer},
  \citenamefont {Izumi}, \citenamefont {Jacobson}, \citenamefont {James},
  \citenamefont {Jang}, \citenamefont {Jang}, \citenamefont {Jesse},
  \citenamefont {Johnson}, \citenamefont {Jones}, \citenamefont {Jones},
  \citenamefont {Jones}, \citenamefont {Ju}, \citenamefont {Kalmus},
  \citenamefont {Kalogera}, \citenamefont {Kandhasamy}, \citenamefont {Kang},
  \citenamefont {Kanner}, \citenamefont {Kasturi}, \citenamefont
  {Katsavounidis}, \citenamefont {Katzman}, \citenamefont {Kaufer},
  \citenamefont {Kawabe}, \citenamefont {Kawamura}, \citenamefont {Kawazoe},
  \citenamefont {Keitel}, \citenamefont {Kelley}, \citenamefont {Kells},
  \citenamefont {Keppel}, \citenamefont {Khalaidovski}, \citenamefont
  {Khalili}, \citenamefont {Khazanov}, \citenamefont {Kim}, \citenamefont
  {Kim}, \citenamefont {Kim}, \citenamefont {Kim}, \citenamefont {Kim},
  \citenamefont {King}, \citenamefont {Kinzel}, \citenamefont {Kissel},
  \citenamefont {Klimenko}, \citenamefont {Kline}, \citenamefont {Kokeyama},
  \citenamefont {Kondrashov}, \citenamefont {Koranda}, \citenamefont {Korth},
  \citenamefont {Kozak}, \citenamefont {Kozameh}, \citenamefont {Kremin},
  \citenamefont {Kringel}, \citenamefont {Krishnan}, \citenamefont
  {Kucharczyk}, \citenamefont {Kuehn}, \citenamefont {Kumar}, \citenamefont
  {Kumar}, \citenamefont {Kuper}, \citenamefont {Kurdyumov}, \citenamefont
  {Kwee}, \citenamefont {Lam}, \citenamefont {Landry}, \citenamefont {Lantz},
  \citenamefont {Lasky}, \citenamefont {Lawrie}, \citenamefont {Lazzarini},
  \citenamefont {Le~Roux}, \citenamefont {Leaci}, \citenamefont {Lee},
  \citenamefont {Lee}, \citenamefont {Lee}, \citenamefont {Lee}, \citenamefont
  {Leong}, \citenamefont {Levine}, \citenamefont {Lhuillier}, \citenamefont
  {Lin}, \citenamefont {Litvine}, \citenamefont {Liu}, \citenamefont {Liu},
  \citenamefont {Lockerbie}, \citenamefont {Lodhia}, \citenamefont {Loew},
  \citenamefont {Logue}, \citenamefont {Lombardi}, \citenamefont {Lormand},
  \citenamefont {Lough}, \citenamefont {Lubinski}, \citenamefont {Lš¹ck},
  \citenamefont {Lundgren}, \citenamefont {Macarthur}, \citenamefont
  {Macdonald}, \citenamefont {Machenschalk}, \citenamefont {MacInnis},
  \citenamefont {Macleod}, \citenamefont {Maga?a-Sandoval}, \citenamefont
  {Mageswaran}, \citenamefont {Mailand}, \citenamefont {Manca}, \citenamefont
  {Mandel}, \citenamefont {Mandic}, \citenamefont {Mš¢rka}, \citenamefont
  {Mš¢rka}, \citenamefont {Markosyan}, \citenamefont {Maros}, \citenamefont
  {Martin}, \citenamefont {Martin}, \citenamefont {Martinov}, \citenamefont
  {Marx}, \citenamefont {Mason}, \citenamefont {Matichard}, \citenamefont
  {Matone}, \citenamefont {Matzner}, \citenamefont {Mavalvala}, \citenamefont
  {May}, \citenamefont {Mazzolo}, \citenamefont {McAuley}, \citenamefont
  {McCarthy}, \citenamefont {McClelland}, \citenamefont {McGuire},
  \citenamefont {McIntyre}, \citenamefont {McIver}, \citenamefont {Meadors},
  \citenamefont {Mehmet}, \citenamefont {Meier}, \citenamefont {Melatos},
  \citenamefont {Mendell}, \citenamefont {Mercer}, \citenamefont {Meshkov},
  \citenamefont {Messenger}, \citenamefont {Meyer}, \citenamefont {Miao},
  \citenamefont {Miller}, \citenamefont {Mingarelli}, \citenamefont {Mitra},
  \citenamefont {Mitrofanov}, \citenamefont {Mitselmakher}, \citenamefont
  {Mittleman}, \citenamefont {Moe}, \citenamefont {Mokler}, \citenamefont
  {Mohapatra}, \citenamefont {Moraru}, \citenamefont {Moreno}, \citenamefont
  {Mori}, \citenamefont {Morriss}, \citenamefont {Mossavi}, \citenamefont
  {Mow-Lowry}, \citenamefont {Mueller}, \citenamefont {Mueller}, \citenamefont
  {Mukherjee}, \citenamefont {Mullavey}, \citenamefont {Munch}, \citenamefont
  {Murphy}, \citenamefont {Murray}, \citenamefont {Mytidis}, \citenamefont
  {Nanda~Kumar}, \citenamefont {Nash}, \citenamefont {Nayak}, \citenamefont
  {Necula}, \citenamefont {Newton}, \citenamefont {Nguyen}, \citenamefont
  {Nishida}, \citenamefont {Nishizawa}, \citenamefont {Nitz}, \citenamefont
  {Nolting}, \citenamefont {Normandin}, \citenamefont {Nuttall}, \citenamefont
  {O'Dell}, \citenamefont {O'Reilly}, \citenamefont {O'Shaughnessy},
  \citenamefont {Ochsner}, \citenamefont {Oelker}, \citenamefont {Ogin},
  \citenamefont {Oh}, \citenamefont {Oh}, \citenamefont {Ohme}, \citenamefont
  {Oppermann}, \citenamefont {Osthelder}, \citenamefont {Ott}, \citenamefont
  {Ottaway}, \citenamefont {Ottens}, \citenamefont {Ou}, \citenamefont
  {Overmier}, \citenamefont {Owen}, \citenamefont {Padilla}, \citenamefont
  {Pai}, \citenamefont {Pan}, \citenamefont {Pankow}, \citenamefont {Papa},
  \citenamefont {Paris}, \citenamefont {Parkinson}, \citenamefont {Pedraza},
  \citenamefont {Penn}, \citenamefont {Peralta}, \citenamefont {Perreca},
  \citenamefont {Phelps}, \citenamefont {Pickenpack}, \citenamefont {Pierro},
  \citenamefont {Pinto}, \citenamefont {Pitkin}, \citenamefont {Pletsch},
  \citenamefont {P?ld}, \citenamefont {Postiglione}, \citenamefont {Poux},
  \citenamefont {Predoi}, \citenamefont {Prestegard}, \citenamefont {Price},
  \citenamefont {Prijatelj}, \citenamefont {Privitera}, \citenamefont
  {Prokhorov}, \citenamefont {Puncken}, \citenamefont {Quetschke},
  \citenamefont {Quintero}, \citenamefont {Quitzow-James}, \citenamefont
  {Raab}, \citenamefont {Radkins}, \citenamefont {Raffai}, \citenamefont
  {Raja}, \citenamefont {Rakhmanov}, \citenamefont {Ramet}, \citenamefont
  {Raymond}, \citenamefont {Reed}, \citenamefont {Reed}, \citenamefont {Reid},
  \citenamefont {Reitze}, \citenamefont {Riesen}, \citenamefont {Riles},
  \citenamefont {Roberts}, \citenamefont {Robertson}, \citenamefont {Robinson},
  \citenamefont {Roddy}, \citenamefont {Rodriguez}, \citenamefont {Rodriguez},
  \citenamefont {Rodruck}, \citenamefont {Rollins}, \citenamefont {Romie},
  \citenamefont {R?ver}, \citenamefont {Rowan}, \citenamefont {Rš¹diger},
  \citenamefont {Ryan}, \citenamefont {Salemi}, \citenamefont {Sammut},
  \citenamefont {Sandberg}, \citenamefont {Sanders}, \citenamefont {Sankar},
  \citenamefont {Sannibale}, \citenamefont {Santamaršªa}, \citenamefont
  {Santiago-Prieto}, \citenamefont {Santostasi}, \citenamefont {Sathyaprakash},
  \citenamefont {Saulson}, \citenamefont {Savage}, \citenamefont {Schilling},
  \citenamefont {Schnabel}, \citenamefont {Schofield}, \citenamefont
  {Schuette}, \citenamefont {Schulz}, \citenamefont {Schutz}, \citenamefont
  {Schwinberg}, \citenamefont {Scott}, \citenamefont {Scott}, \citenamefont
  {Seifert}, \citenamefont {Sellers}, \citenamefont {Sengupta}, \citenamefont
  {Sergeev}, \citenamefont {Shaddock}, \citenamefont {Shahriar}, \citenamefont
  {Shaltev}, \citenamefont {Shao}, \citenamefont {Shapiro}, \citenamefont
  {Shawhan}, \citenamefont {Shoemaker}, \citenamefont {Sidery}, \citenamefont
  {Siemens}, \citenamefont {Sigg}, \citenamefont {Simakov}, \citenamefont
  {Singer}, \citenamefont {Singer}, \citenamefont {Sintes}, \citenamefont
  {Skelton}, \citenamefont {Slagmolen}, \citenamefont {Slutsky}, \citenamefont
  {Smith}, \citenamefont {Smith}, \citenamefont {Smith}, \citenamefont
  {Smith-Lefebvre}, \citenamefont {Son}, \citenamefont {Sorazu}, \citenamefont
  {Souradeep}, \citenamefont {Stefszky}, \citenamefont {Steinert},
  \citenamefont {Steinlechner}, \citenamefont {Steinlechner}, \citenamefont
  {Steplewski}, \citenamefont {Stevens}, \citenamefont {Stochino},
  \citenamefont {Stone}, \citenamefont {Strain}, \citenamefont {Strigin},
  \citenamefont {Stroeer}, \citenamefont {Stuver}, \citenamefont
  {Summerscales}, \citenamefont {Susmithan}, \citenamefont {Sutton},
  \citenamefont {Szeifert}, \citenamefont {Talukder}, \citenamefont {Tanner},
  \citenamefont {Tarabrin}, \citenamefont {Taylor}, \citenamefont {Thomas},
  \citenamefont {Thomas}, \citenamefont {Thorne}, \citenamefont {Thorne},
  \citenamefont {Thrane}, \citenamefont {Tiwari}, \citenamefont {Tokmakov},
  \citenamefont {Tomlinson}, \citenamefont {Torres}, \citenamefont {Torrie},
  \citenamefont {Traylor}, \citenamefont {Tse}, \citenamefont {Ugolini},
  \citenamefont {Unnikrishnan}, \citenamefont {Vahlbruch}, \citenamefont
  {Vallisneri}, \citenamefont {Van Der~Sluys}, \citenamefont {Van~Veggel},
  \citenamefont {Vass}, \citenamefont {Vaulin}, \citenamefont {Vecchio},
  \citenamefont {Veitch}, \citenamefont {Veitch}, \citenamefont {Venkateswara},
  \citenamefont {Verma}, \citenamefont {Vincent-Finley}, \citenamefont
  {Vitale}, \citenamefont {Vo}, \citenamefont {Vorvick}, \citenamefont
  {Vousden}, \citenamefont {Vyatchanin}, \citenamefont {Wade}, \citenamefont
  {Wade}, \citenamefont {Wade}, \citenamefont {Waldman}, \citenamefont
  {Wallace}, \citenamefont {Wan}, \citenamefont {Wang}, \citenamefont {Wang},
  \citenamefont {Wang}, \citenamefont {Wanner}, \citenamefont {Ward},
  \citenamefont {Was}, \citenamefont {Weinert}, \citenamefont {Weinstein},
  \citenamefont {Weiss}, \citenamefont {Welborn}, \citenamefont {Wen},
  \citenamefont {Wessels}, \citenamefont {West}, \citenamefont {Westphal},
  \citenamefont {Wette}, \citenamefont {Whelan}, \citenamefont {Whitcomb},
  \citenamefont {Wiseman}, \citenamefont {White}, \citenamefont {Whiting},
  \citenamefont {Wiesner}, \citenamefont {Wilkinson}, \citenamefont {Willems},
  \citenamefont {Williams}, \citenamefont {Williams}, \citenamefont {Williams},
  \citenamefont {Willis}, \citenamefont {Willke}, \citenamefont {Wimmer},
  \citenamefont {Winkelmann}, \citenamefont {Winkler}, \citenamefont {C.~Wipf},
  \citenamefont {Wittel}, \citenamefont {Woan}, \citenamefont {Wooley},
  \citenamefont {Worden}, \citenamefont {Yablon}, \citenamefont {Yakushin},
  \citenamefont {Yamamoto}, \citenamefont {Yancey}, \citenamefont {Yang},
  \citenamefont {Yeaton-Massey}, \citenamefont {Yoshida}, \citenamefont {Yum},
  \citenamefont {Zanolin}, \citenamefont {Zhang}, \citenamefont {Zhang},
  \citenamefont {Zhao}, \citenamefont {Zhu}, \citenamefont {Zhu}, \citenamefont
  {Zotov}, \citenamefont {Zucker},\ and\ \citenamefont
  {Zweizig}}]{aasi_enhanced_2013}%
  \BibitemOpen
  \bibfield  {author} {\bibinfo {author} {\bibfnamefont {J.}~\bibnamefont
  {Aasi}}, \bibinfo {author} {\bibfnamefont {J.}~\bibnamefont {Abadie}},
  \bibinfo {author} {\bibfnamefont {B.~P.}\ \bibnamefont {Abbott}}, \bibinfo
  {author} {\bibfnamefont {R.}~\bibnamefont {Abbott}}, \bibinfo {author}
  {\bibfnamefont {T.~D.}\ \bibnamefont {Abbott}}, \bibinfo {author}
  {\bibfnamefont {M.~R.}\ \bibnamefont {Abernathy}}, \bibinfo {author}
  {\bibfnamefont {C.}~\bibnamefont {Adams}}, \bibinfo {author} {\bibfnamefont
  {T.}~\bibnamefont {Adams}}, \bibinfo {author} {\bibfnamefont
  {P.}~\bibnamefont {Addesso}}, \bibinfo {author} {\bibfnamefont {R.~X.}\
  \bibnamefont {Adhikari}},~et~al.,\ }\bibfield  {title} {\bibinfo {title}
  {Enhanced sensitivity of the {LIGO} gravitational wave detector by using
  squeezed states of light},\ }\href {\doibase10.1038/nphoton.2013.177}
  {\bibfield  {journal} {\bibinfo  {journal} {Nat. Photon.}\ }\textbf {\bibinfo
  {volume} {7}},\ \bibinfo {pages} {613} (\bibinfo {year} {2013})}\BibitemShut
  {NoStop}%
\bibitem [{\citenamefont {Belenchia}\ \emph {,~et~al.}(2016)\citenamefont
  {Belenchia}, \citenamefont {Benincasa}, \citenamefont {Liberati},
  \citenamefont {Marin}, \citenamefont {Marino},\ and\ \citenamefont
  {Ortolan}}]{belenchia_testing_2016}%
  \BibitemOpen
  \bibfield  {author} {\bibinfo {author} {\bibfnamefont {A.}~\bibnamefont
  {Belenchia}}, \bibinfo {author} {\bibfnamefont {D.~M.}\ \bibnamefont
  {Benincasa}}, \bibinfo {author} {\bibfnamefont {S.}~\bibnamefont {Liberati}},
  \bibinfo {author} {\bibfnamefont {F.}~\bibnamefont {Marin}}, \bibinfo
  {author} {\bibfnamefont {F.}~\bibnamefont {Marino}}, \ and\ \bibinfo {author}
  {\bibfnamefont {A.}~\bibnamefont {Ortolan}},\ }\bibfield  {title} {\bibinfo
  {title} {Testing {Quantum} {Gravity} {Induced} {Nonlocality} via
  {Optomechanical} {Quantum} {Oscillators}},\ }\href
  {\doibase10.1103/PhysRevLett.116.161303} {\bibfield  {journal} {\bibinfo
  {journal} {Phys. Rev. Lett.}\ }\textbf {\bibinfo {volume} {116}},\ \bibinfo
  {pages} {161303} (\bibinfo {year} {2016})}\BibitemShut {NoStop}%
\bibitem [{\citenamefont {Han}\ \emph {,~et~al.}(2021)\citenamefont {Han},
  \citenamefont {Fu}, \citenamefont {Zou}, \citenamefont {Jiang},\ and\
  \citenamefont {Tang}}]{han_microwave-optical_2021}%
  \BibitemOpen
  \bibfield  {author} {\bibinfo {author} {\bibfnamefont {X.}~\bibnamefont
  {Han}}, \bibinfo {author} {\bibfnamefont {W.}~\bibnamefont {Fu}}, \bibinfo
  {author} {\bibfnamefont {C.-L.}\ \bibnamefont {Zou}}, \bibinfo {author}
  {\bibfnamefont {L.}~\bibnamefont {Jiang}}, \ and\ \bibinfo {author}
  {\bibfnamefont {H.~X.}\ \bibnamefont {Tang}},\ }\bibfield  {title} {\bibinfo
  {title} {Microwave-optical quantum frequency conversion},\ }\href
  {\doibase10.1364/OPTICA.425414} {\bibfield  {journal} {\bibinfo  {journal}
  {Optica}\ }\textbf {\bibinfo {volume} {8}},\ \bibinfo {pages} {1050}
  (\bibinfo {year} {2021})}\BibitemShut {NoStop}%
\bibitem [{\citenamefont {Han}\ \emph {,~et~al.}(2020)\citenamefont {Han},
  \citenamefont {Fu}, \citenamefont {Zhong}, \citenamefont {Zou}, \citenamefont
  {Xu}, \citenamefont {Sayem}, \citenamefont {Xu}, \citenamefont {Wang},
  \citenamefont {Cheng}, \citenamefont {Jiang},\ and\ \citenamefont
  {Tang}}]{han_cavity_2020}%
  \BibitemOpen
  \bibfield  {author} {\bibinfo {author} {\bibfnamefont {X.}~\bibnamefont
  {Han}}, \bibinfo {author} {\bibfnamefont {W.}~\bibnamefont {Fu}}, \bibinfo
  {author} {\bibfnamefont {C.}~\bibnamefont {Zhong}}, \bibinfo {author}
  {\bibfnamefont {C.-L.}\ \bibnamefont {Zou}}, \bibinfo {author} {\bibfnamefont
  {Y.}~\bibnamefont {Xu}}, \bibinfo {author} {\bibfnamefont {A.~A.}\
  \bibnamefont {Sayem}}, \bibinfo {author} {\bibfnamefont {M.}~\bibnamefont
  {Xu}}, \bibinfo {author} {\bibfnamefont {S.}~\bibnamefont {Wang}}, \bibinfo
  {author} {\bibfnamefont {R.}~\bibnamefont {Cheng}}, \bibinfo {author}
  {\bibfnamefont {L.}~\bibnamefont {Jiang}}, \ and\ \bibinfo {author}
  {\bibfnamefont {H.~X.}\ \bibnamefont {Tang}},\ }\bibfield  {title} {\bibinfo
  {title} {Cavity piezo-mechanics for superconducting-nanophotonic quantum
  interface},\ }\href {\doibase10.1038/s41467-020-17053-3} {\bibfield
  {journal} {\bibinfo  {journal} {Nat. Commun.}\ }\textbf {\bibinfo {volume}
  {11}},\ \bibinfo {pages} {3237} (\bibinfo {year} {2020})}\BibitemShut
  {NoStop}%
\bibitem [{\citenamefont {Mirhosseini}\ \emph {,~et~al.}(2020)\citenamefont
  {Mirhosseini}, \citenamefont {Sipahigil}, \citenamefont {Kalaee},\ and\
  \citenamefont {Painter}}]{mirhosseini_superconducting_2020}%
  \BibitemOpen
  \bibfield  {author} {\bibinfo {author} {\bibfnamefont {M.}~\bibnamefont
  {Mirhosseini}}, \bibinfo {author} {\bibfnamefont {A.}~\bibnamefont
  {Sipahigil}}, \bibinfo {author} {\bibfnamefont {M.}~\bibnamefont {Kalaee}}, \
  and\ \bibinfo {author} {\bibfnamefont {O.}~\bibnamefont {Painter}},\
  }\bibfield  {title} {\bibinfo {title} {Superconducting qubit to optical
  photon transduction},\ }\href {\doibase10.1038/s41586-020-3038-6} {\bibfield
  {journal} {\bibinfo  {journal} {Nature}\ }\textbf {\bibinfo {volume} {588}},\
  \bibinfo {pages} {599} (\bibinfo {year} {2020})}\BibitemShut {NoStop}%
\bibitem [{\citenamefont {Brubaker}\ \emph {,~et~al.}(2022)\citenamefont
  {Brubaker}, \citenamefont {Kindem}, \citenamefont {Urmey}, \citenamefont
  {Mittal}, \citenamefont {Delaney}, \citenamefont {Burns}, \citenamefont
  {Vissers}, \citenamefont {Lehnert},\ and\ \citenamefont
  {Regal}}]{brubaker_optomechanical_2022}%
  \BibitemOpen
  \bibfield  {author} {\bibinfo {author} {\bibfnamefont {B.}~\bibnamefont
  {Brubaker}}, \bibinfo {author} {\bibfnamefont {J.}~\bibnamefont {Kindem}},
  \bibinfo {author} {\bibfnamefont {M.}~\bibnamefont {Urmey}}, \bibinfo
  {author} {\bibfnamefont {S.}~\bibnamefont {Mittal}}, \bibinfo {author}
  {\bibfnamefont {R.}~\bibnamefont {Delaney}}, \bibinfo {author} {\bibfnamefont
  {P.}~\bibnamefont {Burns}}, \bibinfo {author} {\bibfnamefont
  {M.}~\bibnamefont {Vissers}}, \bibinfo {author} {\bibfnamefont
  {K.}~\bibnamefont {Lehnert}}, \ and\ \bibinfo {author} {\bibfnamefont
  {C.}~\bibnamefont {Regal}},\ }\bibfield  {title} {\bibinfo {title}
  {Optomechanical {Ground}-{State} {Cooling} in a {Continuous} and {Efficient}
  {Electro}-{Optic} {Transducer}},\ }\href {\doibase10.1103/PhysRevX.12.021062}
  {\bibfield  {journal} {\bibinfo  {journal} {Phys. Rev. X}\ }\textbf {\bibinfo
  {volume} {12}},\ \bibinfo {pages} {021062} (\bibinfo {year}
  {2022})}\BibitemShut {NoStop}%
\bibitem [{\citenamefont {Chen}\ \emph {,~et~al.}(2023)\citenamefont {Chen},
  \citenamefont {Li}, \citenamefont {Lee}, \citenamefont {Chakravarthi},
  \citenamefont {Fu},\ and\ \citenamefont {Li}}]{chen_optomechanical_2023}%
  \BibitemOpen
  \bibfield  {author} {\bibinfo {author} {\bibfnamefont {I.-T.}\ \bibnamefont
  {Chen}}, \bibinfo {author} {\bibfnamefont {B.}~\bibnamefont {Li}}, \bibinfo
  {author} {\bibfnamefont {S.}~\bibnamefont {Lee}}, \bibinfo {author}
  {\bibfnamefont {S.}~\bibnamefont {Chakravarthi}}, \bibinfo {author}
  {\bibfnamefont {K.-M.}\ \bibnamefont {Fu}}, \ and\ \bibinfo {author}
  {\bibfnamefont {M.}~\bibnamefont {Li}},\ }\bibfield  {title} {\bibinfo
  {title} {Optomechanical ring resonator for efficient microwave-optical
  frequency conversion},\ }\href {\doibase10.1038/s41467-023-43393-x}
  {\bibfield  {journal} {\bibinfo  {journal} {Nat. Commun.}\ }\textbf {\bibinfo
  {volume} {14}},\ \bibinfo {pages} {7594} (\bibinfo {year}
  {2023})}\BibitemShut {NoStop}%
\bibitem [{\citenamefont {Zhao}\ \emph {,~et~al.}(2025)\citenamefont {Zhao},
  \citenamefont {Chen}, \citenamefont {Kejriwal},\ and\ \citenamefont
  {Mirhosseini}}]{zhao_quantum-enabled_2025}%
  \BibitemOpen
  \bibfield  {author} {\bibinfo {author} {\bibfnamefont {H.}~\bibnamefont
  {Zhao}}, \bibinfo {author} {\bibfnamefont {W.~D.}\ \bibnamefont {Chen}},
  \bibinfo {author} {\bibfnamefont {A.}~\bibnamefont {Kejriwal}}, \ and\
  \bibinfo {author} {\bibfnamefont {M.}~\bibnamefont {Mirhosseini}},\
  }\bibfield  {title} {\bibinfo {title} {Quantum-enabled microwave-to-optical
  transduction via silicon nanomechanics},\ }\href
  {\doibase10.1038/s41565-025-01874-8} {\bibfield  {journal} {\bibinfo
  {journal} {Nat. Nanotech.}\ } (\bibinfo {year} {2025}),\
  10.1038/s41565-025-01874-8}\BibitemShut {NoStop}%
\bibitem [{\citenamefont {Safavi-Naeini}\ \emph {,~et~al.}(2019)\citenamefont
  {Safavi-Naeini}, \citenamefont {Van~Thourhout}, \citenamefont {Baets},\ and\
  \citenamefont {Van~Laer}}]{safavi2019}%
  \BibitemOpen
  \bibfield  {author} {\bibinfo {author} {\bibfnamefont {A.~H.}\ \bibnamefont
  {Safavi-Naeini}}, \bibinfo {author} {\bibfnamefont {D.}~\bibnamefont
  {Van~Thourhout}}, \bibinfo {author} {\bibfnamefont {R.}~\bibnamefont
  {Baets}}, \ and\ \bibinfo {author} {\bibfnamefont {R.}~\bibnamefont
  {Van~Laer}},\ }\bibfield  {title} {\bibinfo {title} {Controlling phonons and
  photons at the wavelength scale: integrated photonics meets integrated
  phononics},\ }\href {\doibase10.1364/OPTICA.6.000213} {\bibfield  {journal}
  {\bibinfo  {journal} {Optica}\ }\textbf {\bibinfo {volume} {6}},\ \bibinfo
  {pages} {213} (\bibinfo {year} {2019})}\BibitemShut {NoStop}%
\bibitem [{\citenamefont {Clerk}\ \emph {,~et~al.}(2020)\citenamefont {Clerk},
  \citenamefont {Lehnert}, \citenamefont {Bertet}, \citenamefont {Petta},\ and\
  \citenamefont {Nakamura}}]{Clerk2020}%
  \BibitemOpen
  \bibfield  {author} {\bibinfo {author} {\bibfnamefont {A.~A.}\ \bibnamefont
  {Clerk}}, \bibinfo {author} {\bibfnamefont {K.~W.}\ \bibnamefont {Lehnert}},
  \bibinfo {author} {\bibfnamefont {P.}~\bibnamefont {Bertet}}, \bibinfo
  {author} {\bibfnamefont {J.~R.}\ \bibnamefont {Petta}}, \ and\ \bibinfo
  {author} {\bibfnamefont {Y.}~\bibnamefont {Nakamura}},\ }\bibfield  {title}
  {\bibinfo {title} {{Hybrid quantum systems with circuit quantum
  electrodynamics}},\ }\href {\doibase10.1038/s41567-020-0797-9} {\bibfield
  {journal} {\bibinfo  {journal} {Nature Physics}\ }\textbf {\bibinfo {volume}
  {16}},\ \bibinfo {pages} {257} (\bibinfo {year} {2020})}\BibitemShut
  {NoStop}%
\bibitem [{\citenamefont {Barzanjeh}\ \emph
  {,~et~al.}(2022{\natexlab{b}})\citenamefont {Barzanjeh}, \citenamefont
  {Xuereb}, \citenamefont {Gr{\"{o}}blacher}, \citenamefont {Paternostro},
  \citenamefont {Regal},\ and\ \citenamefont {Weig}}]{Barzanjeh2022}%
  \BibitemOpen
  \bibfield  {author} {\bibinfo {author} {\bibfnamefont {S.}~\bibnamefont
  {Barzanjeh}}, \bibinfo {author} {\bibfnamefont {A.}~\bibnamefont {Xuereb}},
  \bibinfo {author} {\bibfnamefont {S.}~\bibnamefont {Gr{\"{o}}blacher}},
  \bibinfo {author} {\bibfnamefont {M.}~\bibnamefont {Paternostro}}, \bibinfo
  {author} {\bibfnamefont {C.~A.}\ \bibnamefont {Regal}}, \ and\ \bibinfo
  {author} {\bibfnamefont {E.~M.}\ \bibnamefont {Weig}},\ }\bibfield  {title}
  {\bibinfo {title} {{Optomechanics for quantum technologies}},\ }\href
  {\doibase10.1038/s41567-021-01402-0} {\bibfield  {journal} {\bibinfo
  {journal} {Nature Physics}\ }\textbf {\bibinfo {volume} {18}},\ \bibinfo
  {pages} {15} (\bibinfo {year} {2022}{\natexlab{b}})}\BibitemShut {NoStop}%
\bibitem [{\citenamefont {Chan}\ \emph {,~et~al.}(2011)\citenamefont {Chan},
  \citenamefont {Alegre}, \citenamefont {Safavi-Naeini}, \citenamefont {Hill},
  \citenamefont {Krause}, \citenamefont {Groblacher}, \citenamefont
  {Aspelmeyer},\ and\ \citenamefont {Painter}}]{chan_laser_2011}%
  \BibitemOpen
  \bibfield  {author} {\bibinfo {author} {\bibfnamefont {J.}~\bibnamefont
  {Chan}}, \bibinfo {author} {\bibfnamefont {T.~P.~M.}\ \bibnamefont {Alegre}},
  \bibinfo {author} {\bibfnamefont {A.~H.}\ \bibnamefont {Safavi-Naeini}},
  \bibinfo {author} {\bibfnamefont {J.~T.}\ \bibnamefont {Hill}}, \bibinfo
  {author} {\bibfnamefont {A.}~\bibnamefont {Krause}}, \bibinfo {author}
  {\bibfnamefont {S.}~\bibnamefont {Groblacher}}, \bibinfo {author}
  {\bibfnamefont {M.}~\bibnamefont {Aspelmeyer}}, \ and\ \bibinfo {author}
  {\bibfnamefont {O.}~\bibnamefont {Painter}},\ }\bibfield  {title} {\bibinfo
  {title} {Laser cooling of a nanomechanical oscillator into its quantum ground
  state},\ }\href {\doibase10.1038/nature10461} {\bibfield  {journal} {\bibinfo
   {journal} {Nature}\ }\textbf {\bibinfo {volume} {478}},\ \bibinfo {pages}
  {89} (\bibinfo {year} {2011})}\BibitemShut {NoStop}%
\bibitem [{\citenamefont {Fan}\ \emph {,~et~al.}(2016)\citenamefont {Fan},
  \citenamefont {Zou}, \citenamefont {Poot}, \citenamefont {Cheng},
  \citenamefont {Guo}, \citenamefont {Han},\ and\ \citenamefont
  {Tang}}]{fan_integrated_2016}%
  \BibitemOpen
  \bibfield  {author} {\bibinfo {author} {\bibfnamefont {L.}~\bibnamefont
  {Fan}}, \bibinfo {author} {\bibfnamefont {C.-L.}\ \bibnamefont {Zou}},
  \bibinfo {author} {\bibfnamefont {M.}~\bibnamefont {Poot}}, \bibinfo {author}
  {\bibfnamefont {R.}~\bibnamefont {Cheng}}, \bibinfo {author} {\bibfnamefont
  {X.}~\bibnamefont {Guo}}, \bibinfo {author} {\bibfnamefont {X.}~\bibnamefont
  {Han}}, \ and\ \bibinfo {author} {\bibfnamefont {H.~X.}\ \bibnamefont
  {Tang}},\ }\bibfield  {title} {\bibinfo {title} {Integrated optomechanical
  single-photon frequency shifter},\ }\href {\doibase10.1038/nphoton.2016.206}
  {\bibfield  {journal} {\bibinfo  {journal} {Nat. Photon.}\ }\textbf {\bibinfo
  {volume} {10}},\ \bibinfo {pages} {766} (\bibinfo {year} {2016})}\BibitemShut
  {NoStop}%
\bibitem [{\citenamefont {Otterstrom}\ \emph {,~et~al.}(2018)\citenamefont
  {Otterstrom}, \citenamefont {Behunin}, \citenamefont {Kittlaus},
  \citenamefont {Wang},\ and\ \citenamefont
  {Rakich}}]{otterstrom_silicon_2018}%
  \BibitemOpen
  \bibfield  {author} {\bibinfo {author} {\bibfnamefont {N.~T.}\ \bibnamefont
  {Otterstrom}}, \bibinfo {author} {\bibfnamefont {R.~O.}\ \bibnamefont
  {Behunin}}, \bibinfo {author} {\bibfnamefont {E.~A.}\ \bibnamefont
  {Kittlaus}}, \bibinfo {author} {\bibfnamefont {Z.}~\bibnamefont {Wang}}, \
  and\ \bibinfo {author} {\bibfnamefont {P.~T.}\ \bibnamefont {Rakich}},\
  }\bibfield  {title} {\bibinfo {title} {A silicon {Brillouin} laser},\ }\href
  {\doibase10.1126/science.aar6113} {\bibfield  {journal} {\bibinfo  {journal}
  {Science}\ }\textbf {\bibinfo {volume} {360}},\ \bibinfo {pages} {1113}
  (\bibinfo {year} {2018})}\BibitemShut {NoStop}%
\bibitem [{\citenamefont {Yu}\ \emph {,~et~al.}(2025)\citenamefont {Yu},
  \citenamefont {Zhou}, \citenamefont {Yang}, \citenamefont {Zhang},
  \citenamefont {Zhu}, \citenamefont {Yang}, \citenamefont {Xu}, \citenamefont
  {Chen}, \citenamefont {Zou},\ and\ \citenamefont {Lu}}]{yu-chip-2025}%
  \BibitemOpen
  \bibfield  {author} {\bibinfo {author} {\bibfnamefont {S.}~\bibnamefont
  {Yu}}, \bibinfo {author} {\bibfnamefont {R.}~\bibnamefont {Zhou}}, \bibinfo
  {author} {\bibfnamefont {G.}~\bibnamefont {Yang}}, \bibinfo {author}
  {\bibfnamefont {Q.}~\bibnamefont {Zhang}}, \bibinfo {author} {\bibfnamefont
  {H.}~\bibnamefont {Zhu}}, \bibinfo {author} {\bibfnamefont {Y.}~\bibnamefont
  {Yang}}, \bibinfo {author} {\bibfnamefont {X.}~\bibnamefont {Xu}}, \bibinfo
  {author} {\bibfnamefont {B.}~\bibnamefont {Chen}}, \bibinfo {author}
  {\bibfnamefont {C.}~\bibnamefont {Zou}}, \ and\ \bibinfo {author}
  {\bibfnamefont {J.}~\bibnamefont {Lu}},\ }\bibfield  {title} {\bibinfo
  {title} {On-chip {Brillouin} {Amplifier} in {Suspended} {Lithium} {Niobate}
  {Nanowaveguides}},\ }\href {\doibase10.1002/lpor.202500027} {\bibfield
  {journal} {\bibinfo  {journal} {Laser Photonics Rev.}\ }\textbf {\bibinfo
  {volume} {19}},\ \bibinfo {pages} {2500027} (\bibinfo {year}
  {2025})}\BibitemShut {NoStop}%
\bibitem [{\citenamefont {Xu}\ \emph
  {,~et~al.}(2022{\natexlab{a}})\citenamefont {Xu}, \citenamefont {Wang},
  \citenamefont {Sun},\ and\ \citenamefont {Zou}}]{Xu2022}%
  \BibitemOpen
  \bibfield  {author} {\bibinfo {author} {\bibfnamefont {X.-B.}\ \bibnamefont
  {Xu}}, \bibinfo {author} {\bibfnamefont {W.-T.}\ \bibnamefont {Wang}},
  \bibinfo {author} {\bibfnamefont {L.-Y.}\ \bibnamefont {Sun}}, \ and\
  \bibinfo {author} {\bibfnamefont {C.-L.}\ \bibnamefont {Zou}},\ }\bibfield
  {title} {\bibinfo {title} {{Hybrid superconducting photonic-phononic chip for
  quantum information processing}},\ }\href
  {\doibase10.1016/j.chip.2022.100016} {\bibfield  {journal} {\bibinfo
  {journal} {Chip}\ }\textbf {\bibinfo {volume} {1}},\ \bibinfo {pages}
  {100016} (\bibinfo {year} {2022}{\natexlab{a}})}\BibitemShut {NoStop}%
\bibitem [{\citenamefont {Eggleton}\ \emph {,~et~al.}(2019)\citenamefont
  {Eggleton}, \citenamefont {Poulton}, \citenamefont {Rakich}, \citenamefont
  {Steel},\ and\ \citenamefont {Bahl}}]{eggleton_brillouin_2019}%
  \BibitemOpen
  \bibfield  {author} {\bibinfo {author} {\bibfnamefont {B.~J.}\ \bibnamefont
  {Eggleton}}, \bibinfo {author} {\bibfnamefont {C.~G.}\ \bibnamefont
  {Poulton}}, \bibinfo {author} {\bibfnamefont {P.~T.}\ \bibnamefont {Rakich}},
  \bibinfo {author} {\bibfnamefont {M.~J.}\ \bibnamefont {Steel}}, \ and\
  \bibinfo {author} {\bibfnamefont {G.}~\bibnamefont {Bahl}},\ }\bibfield
  {title} {\bibinfo {title} {Brillouin integrated photonics},\ }\href
  {\doibase10.1038/s41566-019-0498-z} {\bibfield  {journal} {\bibinfo
  {journal} {Nat. Photon.}\ }\textbf {\bibinfo {volume} {13}},\ \bibinfo
  {pages} {664} (\bibinfo {year} {2019})}\BibitemShut {NoStop}%
\bibitem [{\citenamefont {Merklein}\ \emph {,~et~al.}(2022)\citenamefont
  {Merklein}, \citenamefont {Kabakova}, \citenamefont {Zarifi},\ and\
  \citenamefont {Eggleton}}]{merklein_100_2022}%
  \BibitemOpen
  \bibfield  {author} {\bibinfo {author} {\bibfnamefont {M.}~\bibnamefont
  {Merklein}}, \bibinfo {author} {\bibfnamefont {I.~V.}\ \bibnamefont
  {Kabakova}}, \bibinfo {author} {\bibfnamefont {A.}~\bibnamefont {Zarifi}}, \
  and\ \bibinfo {author} {\bibfnamefont {B.~J.}\ \bibnamefont {Eggleton}},\
  }\bibfield  {title} {\bibinfo {title} {100 years of {Brillouin} scattering:
  {Historical} and future perspectives},\ }\href {\doibase10.1063/5.0095488}
  {\bibfield  {journal} {\bibinfo  {journal} {Appl. Phys. Rev.}\ }\textbf
  {\bibinfo {volume} {9}},\ \bibinfo {pages} {041306} (\bibinfo {year}
  {2022})}\BibitemShut {NoStop}%
\bibitem [{\citenamefont {Morrison}\ \emph {,~et~al.}(2017)\citenamefont
  {Morrison}, \citenamefont {Casas-Bedoya}, \citenamefont {Ren}, \citenamefont
  {Vu}, \citenamefont {Liu}, \citenamefont {Zarifi}, \citenamefont {Nguyen},
  \citenamefont {Choi}, \citenamefont {Marpaung}, \citenamefont {Madden},
  \citenamefont {Mitchell},\ and\ \citenamefont
  {Eggleton}}]{morrison_compact_2017}%
  \BibitemOpen
  \bibfield  {author} {\bibinfo {author} {\bibfnamefont {B.}~\bibnamefont
  {Morrison}}, \bibinfo {author} {\bibfnamefont {A.}~\bibnamefont
  {Casas-Bedoya}}, \bibinfo {author} {\bibfnamefont {G.}~\bibnamefont {Ren}},
  \bibinfo {author} {\bibfnamefont {K.}~\bibnamefont {Vu}}, \bibinfo {author}
  {\bibfnamefont {Y.}~\bibnamefont {Liu}}, \bibinfo {author} {\bibfnamefont
  {A.}~\bibnamefont {Zarifi}}, \bibinfo {author} {\bibfnamefont {T.~G.}\
  \bibnamefont {Nguyen}}, \bibinfo {author} {\bibfnamefont {D.-Y.}\
  \bibnamefont {Choi}}, \bibinfo {author} {\bibfnamefont {D.}~\bibnamefont
  {Marpaung}}, \bibinfo {author} {\bibfnamefont {S.~J.}\ \bibnamefont
  {Madden}},~et~al.,\ }\bibfield  {title} {\bibinfo {title} {Compact
  {Brillouin} devices through hybrid integration on silicon},\ }\href
  {\doibase10.1364/OPTICA.4.000847} {\bibfield  {journal} {\bibinfo  {journal}
  {Optica}\ }\textbf {\bibinfo {volume} {4}},\ \bibinfo {pages} {847} (\bibinfo
  {year} {2017})}\BibitemShut {NoStop}%
\bibitem [{\citenamefont {Gundavarapu}\ \emph {,~et~al.}(2019)\citenamefont
  {Gundavarapu}, \citenamefont {Brodnik}, \citenamefont {Puckett},
  \citenamefont {Huffman}, \citenamefont {Bose}, \citenamefont {Behunin},
  \citenamefont {Wu}, \citenamefont {Qiu}, \citenamefont {Pinho}, \citenamefont
  {Chauhan}, \citenamefont {Nohava}, \citenamefont {Rakich}, \citenamefont
  {Nelson}, \citenamefont {Salit},\ and\ \citenamefont
  {Blumenthal}}]{gundavarapu_sub-hertz_2019}%
  \BibitemOpen
  \bibfield  {author} {\bibinfo {author} {\bibfnamefont {S.}~\bibnamefont
  {Gundavarapu}}, \bibinfo {author} {\bibfnamefont {G.~M.}\ \bibnamefont
  {Brodnik}}, \bibinfo {author} {\bibfnamefont {M.}~\bibnamefont {Puckett}},
  \bibinfo {author} {\bibfnamefont {T.}~\bibnamefont {Huffman}}, \bibinfo
  {author} {\bibfnamefont {D.}~\bibnamefont {Bose}}, \bibinfo {author}
  {\bibfnamefont {R.}~\bibnamefont {Behunin}}, \bibinfo {author} {\bibfnamefont
  {J.}~\bibnamefont {Wu}}, \bibinfo {author} {\bibfnamefont {T.}~\bibnamefont
  {Qiu}}, \bibinfo {author} {\bibfnamefont {C.}~\bibnamefont {Pinho}}, \bibinfo
  {author} {\bibfnamefont {N.}~\bibnamefont {Chauhan}},~et~al.,\ }\bibfield
  {title} {\bibinfo {title} {Sub-hertz fundamental linewidth photonic
  integrated {Brillouin} laser},\ }\href {\doibase10.1038/s41566-018-0313-2}
  {\bibfield  {journal} {\bibinfo  {journal} {Nat. Photon.}\ }\textbf {\bibinfo
  {volume} {13}},\ \bibinfo {pages} {60} (\bibinfo {year} {2019})}\BibitemShut
  {NoStop}%
\bibitem [{\citenamefont {Li}\ \emph {,~et~al.}(2013)\citenamefont {Li},
  \citenamefont {Lee},\ and\ \citenamefont {Vahala}}]{li_microwave_2013}%
  \BibitemOpen
  \bibfield  {author} {\bibinfo {author} {\bibfnamefont {J.}~\bibnamefont
  {Li}}, \bibinfo {author} {\bibfnamefont {H.}~\bibnamefont {Lee}}, \ and\
  \bibinfo {author} {\bibfnamefont {K.~J.}\ \bibnamefont {Vahala}},\ }\bibfield
   {title} {\bibinfo {title} {Microwave synthesizer using an on-chip
  {Brillouin} oscillator},\ }\href {\doibase10.1038/ncomms3097} {\bibfield
  {journal} {\bibinfo  {journal} {Nat. Commun.}\ }\textbf {\bibinfo {volume}
  {4}},\ \bibinfo {pages} {2097} (\bibinfo {year} {2013})}\BibitemShut
  {NoStop}%
\bibitem [{\citenamefont {Marpaung}\ \emph {,~et~al.}(2015)\citenamefont
  {Marpaung}, \citenamefont {Morrison}, \citenamefont {Pagani}, \citenamefont
  {Pant}, \citenamefont {Choi}, \citenamefont {Luther-Davies}, \citenamefont
  {Madden},\ and\ \citenamefont {Eggleton}}]{marpaung_low-power_2015}%
  \BibitemOpen
  \bibfield  {author} {\bibinfo {author} {\bibfnamefont {D.}~\bibnamefont
  {Marpaung}}, \bibinfo {author} {\bibfnamefont {B.}~\bibnamefont {Morrison}},
  \bibinfo {author} {\bibfnamefont {M.}~\bibnamefont {Pagani}}, \bibinfo
  {author} {\bibfnamefont {R.}~\bibnamefont {Pant}}, \bibinfo {author}
  {\bibfnamefont {D.-Y.}\ \bibnamefont {Choi}}, \bibinfo {author}
  {\bibfnamefont {B.}~\bibnamefont {Luther-Davies}}, \bibinfo {author}
  {\bibfnamefont {S.~J.}\ \bibnamefont {Madden}}, \ and\ \bibinfo {author}
  {\bibfnamefont {B.~J.}\ \bibnamefont {Eggleton}},\ }\bibfield  {title}
  {\bibinfo {title} {Low-power, chip-based stimulated {Brillouin} scattering
  microwave photonic filter with ultrahigh selectivity},\ }\href
  {\doibase10.1364/OPTICA.2.000076} {\bibfield  {journal} {\bibinfo  {journal}
  {Optica}\ }\textbf {\bibinfo {volume} {2}},\ \bibinfo {pages} {76} (\bibinfo
  {year} {2015})}\BibitemShut {NoStop}%
\bibitem [{\citenamefont {Ye}\ \emph
  {,~et~al.}(2025{\natexlab{a}})\citenamefont {Ye}, \citenamefont {Feng},
  \citenamefont {te~Morsche}, \citenamefont {Wei}, \citenamefont {Klaver},
  \citenamefont {Mishra}, \citenamefont {Zheng}, \citenamefont {Keloth},
  \citenamefont {Tarik~Isik}, \citenamefont {Chen}, \citenamefont {Wang},\ and\
  \citenamefont {Marpaung}}]{ye_integrated_2025}%
  \BibitemOpen
  \bibfield  {author} {\bibinfo {author} {\bibfnamefont {K.}~\bibnamefont
  {Ye}}, \bibinfo {author} {\bibfnamefont {H.}~\bibnamefont {Feng}}, \bibinfo
  {author} {\bibfnamefont {R.}~\bibnamefont {te~Morsche}}, \bibinfo {author}
  {\bibfnamefont {C.}~\bibnamefont {Wei}}, \bibinfo {author} {\bibfnamefont
  {Y.}~\bibnamefont {Klaver}}, \bibinfo {author} {\bibfnamefont
  {A.}~\bibnamefont {Mishra}}, \bibinfo {author} {\bibfnamefont
  {Z.}~\bibnamefont {Zheng}}, \bibinfo {author} {\bibfnamefont
  {A.}~\bibnamefont {Keloth}}, \bibinfo {author} {\bibfnamefont
  {A.}~\bibnamefont {Tarik~Isik}}, \bibinfo {author} {\bibfnamefont
  {Z.}~\bibnamefont {Chen}},~et~al.,\ }\bibfield  {title} {\bibinfo {title}
  {Integrated {Brillouin} photonics in thin-film lithium niobate},\ }\href
  {\doibase10.1126/sciadv.adv4022} {\bibfield  {journal} {\bibinfo  {journal}
  {Sci. Adv.}\ }\textbf {\bibinfo {volume} {11}},\ \bibinfo {pages} {eadv4022}
  (\bibinfo {year} {2025}{\natexlab{a}})}\BibitemShut {NoStop}%
\bibitem [{\citenamefont {Dong}\ \emph {,~et~al.}(2015)\citenamefont {Dong},
  \citenamefont {Shen}, \citenamefont {Zou}, \citenamefont {Zhang},
  \citenamefont {Fu},\ and\ \citenamefont {Guo}}]{dong2015}%
  \BibitemOpen
  \bibfield  {author} {\bibinfo {author} {\bibfnamefont {C.-H.}\ \bibnamefont
  {Dong}}, \bibinfo {author} {\bibfnamefont {Z.}~\bibnamefont {Shen}}, \bibinfo
  {author} {\bibfnamefont {C.-L.}\ \bibnamefont {Zou}}, \bibinfo {author}
  {\bibfnamefont {Y.-L.}\ \bibnamefont {Zhang}}, \bibinfo {author}
  {\bibfnamefont {W.}~\bibnamefont {Fu}}, \ and\ \bibinfo {author}
  {\bibfnamefont {G.-C.}\ \bibnamefont {Guo}},\ }\bibfield  {title} {\bibinfo
  {title} {Brillouin-scattering-induced transparency and non-reciprocal light
  storage},\ }\href {\doibase10.1038/ncomms7193} {\bibfield  {journal}
  {\bibinfo  {journal} {Nat. Commun.}\ }\textbf {\bibinfo {volume} {6}},\
  \bibinfo {pages} {6193} (\bibinfo {year} {2015})}\BibitemShut {NoStop}%
\bibitem [{\citenamefont {Kittlaus}\ \emph {,~et~al.}(2018)\citenamefont
  {Kittlaus}, \citenamefont {Otterstrom}, \citenamefont {Kharel}, \citenamefont
  {Gertler},\ and\ \citenamefont {Rakich}}]{kittlaus2018non}%
  \BibitemOpen
  \bibfield  {author} {\bibinfo {author} {\bibfnamefont {E.~A.}\ \bibnamefont
  {Kittlaus}}, \bibinfo {author} {\bibfnamefont {N.~T.}\ \bibnamefont
  {Otterstrom}}, \bibinfo {author} {\bibfnamefont {P.}~\bibnamefont {Kharel}},
  \bibinfo {author} {\bibfnamefont {S.}~\bibnamefont {Gertler}}, \ and\
  \bibinfo {author} {\bibfnamefont {P.~T.}\ \bibnamefont {Rakich}},\ }\bibfield
   {title} {\bibinfo {title} {Non-reciprocal interband brillouin modulation},\
  }\href {\doibase10.1038/s41566-018-0254-9} {\bibfield  {journal} {\bibinfo
  {journal} {Nat. Photon.}\ }\textbf {\bibinfo {volume} {12}},\ \bibinfo
  {pages} {613} (\bibinfo {year} {2018})}\BibitemShut {NoStop}%
\bibitem [{\citenamefont {Van~Laer}\ \emph {,~et~al.}(2016)\citenamefont
  {Van~Laer}, \citenamefont {Baets},\ and\ \citenamefont
  {Van~Thourhout}}]{van2016}%
  \BibitemOpen
  \bibfield  {author} {\bibinfo {author} {\bibfnamefont {R.}~\bibnamefont
  {Van~Laer}}, \bibinfo {author} {\bibfnamefont {R.}~\bibnamefont {Baets}}, \
  and\ \bibinfo {author} {\bibfnamefont {D.}~\bibnamefont {Van~Thourhout}},\
  }\bibfield  {title} {\bibinfo {title} {Unifying {Brillouin} scattering and
  cavity optomechanics},\ }\href {\doibase10.1103/PhysRevA.93.053828}
  {\bibfield  {journal} {\bibinfo  {journal} {Phys. Rev. A}\ }\textbf {\bibinfo
  {volume} {93}},\ \bibinfo {pages} {053828} (\bibinfo {year}
  {2016})}\BibitemShut {NoStop}%
\bibitem [{\citenamefont {Botter}\ \emph {,~et~al.}(2022)\citenamefont
  {Botter}, \citenamefont {Ye}, \citenamefont {Klaver}, \citenamefont
  {Suryadharma}, \citenamefont {Daulay}, \citenamefont {Liu}, \citenamefont
  {van~den Hoogen}, \citenamefont {Kanger}, \citenamefont {van~der Slot},
  \citenamefont {Klein}, \citenamefont {Hoekman}, \citenamefont {Roeloffzen},
  \citenamefont {Liu},\ and\ \citenamefont {Marpaung}}]{Botter2022}%
  \BibitemOpen
  \bibfield  {author} {\bibinfo {author} {\bibfnamefont {R.}~\bibnamefont
  {Botter}}, \bibinfo {author} {\bibfnamefont {K.}~\bibnamefont {Ye}}, \bibinfo
  {author} {\bibfnamefont {Y.}~\bibnamefont {Klaver}}, \bibinfo {author}
  {\bibfnamefont {R.}~\bibnamefont {Suryadharma}}, \bibinfo {author}
  {\bibfnamefont {O.}~\bibnamefont {Daulay}}, \bibinfo {author} {\bibfnamefont
  {G.}~\bibnamefont {Liu}}, \bibinfo {author} {\bibfnamefont {J.}~\bibnamefont
  {van~den Hoogen}}, \bibinfo {author} {\bibfnamefont {L.}~\bibnamefont
  {Kanger}}, \bibinfo {author} {\bibfnamefont {P.}~\bibnamefont {van~der
  Slot}}, \bibinfo {author} {\bibfnamefont {E.}~\bibnamefont {Klein}},~et~al.,\
  }\bibfield  {title} {\bibinfo {title} {{Guided-acoustic stimulated Brillouin
  scattering in silicon nitride photonic circuits}},\ }\href
  {\doibase10.1126/sciadv.abq2196} {\bibfield  {journal} {\bibinfo  {journal}
  {Science Advances}\ }\textbf {\bibinfo {volume} {8}},\ \bibinfo {pages}
  {abq2196} (\bibinfo {year} {2022})}\BibitemShut {NoStop}%
\bibitem [{\citenamefont {Klaver}\ \emph {,~et~al.}(2024)\citenamefont
  {Klaver}, \citenamefont {te~Morsche}, \citenamefont {Botter}, \citenamefont
  {Hashemi}, \citenamefont {Frare}, \citenamefont {Mishra}, \citenamefont {Ye},
  \citenamefont {Mbonde}, \citenamefont {Ahmadi}, \citenamefont {Taleghani},
  \citenamefont {Jonker}, \citenamefont {Braamhaar}, \citenamefont
  {Selvaganapathy}, \citenamefont {Mascher}, \citenamefont {van~der Slot},
  \citenamefont {Bradley},\ and\ \citenamefont {Marpaung}}]{Klaver2024}%
  \BibitemOpen
  \bibfield  {author} {\bibinfo {author} {\bibfnamefont {Y.}~\bibnamefont
  {Klaver}}, \bibinfo {author} {\bibfnamefont {R.}~\bibnamefont {te~Morsche}},
  \bibinfo {author} {\bibfnamefont {R.~A.}\ \bibnamefont {Botter}}, \bibinfo
  {author} {\bibfnamefont {B.}~\bibnamefont {Hashemi}}, \bibinfo {author}
  {\bibfnamefont {B.~L.~S.}\ \bibnamefont {Frare}}, \bibinfo {author}
  {\bibfnamefont {A.}~\bibnamefont {Mishra}}, \bibinfo {author} {\bibfnamefont
  {K.}~\bibnamefont {Ye}}, \bibinfo {author} {\bibfnamefont {H.}~\bibnamefont
  {Mbonde}}, \bibinfo {author} {\bibfnamefont {P.~T.}\ \bibnamefont {Ahmadi}},
  \bibinfo {author} {\bibfnamefont {N.~M.}\ \bibnamefont {Taleghani}},~et~al.,\
  }\bibfield  {title} {\bibinfo {title} {{Surface acoustic waves Brillouin
  photonics on a silicon nitride chip}},\ }\href
  {http://arxiv.org/abs/2410.16263} {\bibfield  {journal} {\bibinfo  {journal}
  {arXiv preprint: 2410.16263}\ } (\bibinfo {year} {2024})}\BibitemShut
  {NoStop}%
\bibitem [{\citenamefont {Neijts}\ \emph {,~et~al.}(2024)\citenamefont
  {Neijts}, \citenamefont {Lai}, \citenamefont {Riseng}, \citenamefont {Choi},
  \citenamefont {Yan}, \citenamefont {Marpaung}, \citenamefont {Madden},
  \citenamefont {Eggleton},\ and\ \citenamefont {Merklein}}]{Neijts2024}%
  \BibitemOpen
  \bibfield  {author} {\bibinfo {author} {\bibfnamefont {G.}~\bibnamefont
  {Neijts}}, \bibinfo {author} {\bibfnamefont {C.~K.}\ \bibnamefont {Lai}},
  \bibinfo {author} {\bibfnamefont {M.~K.}\ \bibnamefont {Riseng}}, \bibinfo
  {author} {\bibfnamefont {D.-Y.}\ \bibnamefont {Choi}}, \bibinfo {author}
  {\bibfnamefont {K.}~\bibnamefont {Yan}}, \bibinfo {author} {\bibfnamefont
  {D.}~\bibnamefont {Marpaung}}, \bibinfo {author} {\bibfnamefont {S.~J.}\
  \bibnamefont {Madden}}, \bibinfo {author} {\bibfnamefont {B.~J.}\
  \bibnamefont {Eggleton}}, \ and\ \bibinfo {author} {\bibfnamefont
  {M.}~\bibnamefont {Merklein}},\ }\bibfield  {title} {\bibinfo {title}
  {{On-chip stimulated Brillouin scattering via surface acoustic waves}},\
  }\href {\doibase10.1063/5.0220496} {\bibfield  {journal} {\bibinfo  {journal}
  {APL Photonics}\ }\textbf {\bibinfo {volume} {9}},\ \bibinfo {pages} {106114}
  (\bibinfo {year} {2024})}\BibitemShut {NoStop}%
\bibitem [{\citenamefont {Ye}\ \emph
  {,~et~al.}(2025{\natexlab{b}})\citenamefont {Ye}, \citenamefont {Keloth},
  \citenamefont {Marin}, \citenamefont {Cherchi}, \citenamefont {Aalto},\ and\
  \citenamefont {Marpaung}}]{Ye2025}%
  \BibitemOpen
  \bibfield  {author} {\bibinfo {author} {\bibfnamefont {K.}~\bibnamefont
  {Ye}}, \bibinfo {author} {\bibfnamefont {A.}~\bibnamefont {Keloth}}, \bibinfo
  {author} {\bibfnamefont {Y.~E.}\ \bibnamefont {Marin}}, \bibinfo {author}
  {\bibfnamefont {M.}~\bibnamefont {Cherchi}}, \bibinfo {author} {\bibfnamefont
  {T.}~\bibnamefont {Aalto}}, \ and\ \bibinfo {author} {\bibfnamefont
  {D.}~\bibnamefont {Marpaung}},\ }\bibfield  {title} {\bibinfo {title}
  {{Stimulated Brillouin scattering in a non-suspended ultra-low-loss thick-SOI
  platform}},\ }\href {\doibase10.1063/5.0246281} {\bibfield  {journal}
  {\bibinfo  {journal} {APL Photonics}\ }\textbf {\bibinfo {volume} {10}},\
  \bibinfo {pages} {026108} (\bibinfo {year} {2025}{\natexlab{b}})}\BibitemShut
  {NoStop}%
\bibitem [{\citenamefont {Rodrigues}\ \emph {,~et~al.}(2025)\citenamefont
  {Rodrigues}, \citenamefont {Schilder}, \citenamefont {Zurita}, \citenamefont
  {Magalhaes}, \citenamefont {Shams-Ansari}, \citenamefont {dos Santos},
  \citenamefont {Paiano}, \citenamefont {Alegre}, \citenamefont {Loncar},\ and\
  \citenamefont {Wiederhecker}}]{rodrigues_cross-polarized_2025}%
  \BibitemOpen
  \bibfield  {author} {\bibinfo {author} {\bibfnamefont {C.~C.}\ \bibnamefont
  {Rodrigues}}, \bibinfo {author} {\bibfnamefont {N.~J.}\ \bibnamefont
  {Schilder}}, \bibinfo {author} {\bibfnamefont {R.~O.}\ \bibnamefont
  {Zurita}}, \bibinfo {author} {\bibfnamefont {L.~S.}\ \bibnamefont
  {Magalhaes}}, \bibinfo {author} {\bibfnamefont {A.}~\bibnamefont
  {Shams-Ansari}}, \bibinfo {author} {\bibfnamefont {F.~J.}\ \bibnamefont {dos
  Santos}}, \bibinfo {author} {\bibfnamefont {O.~M.}\ \bibnamefont {Paiano}},
  \bibinfo {author} {\bibfnamefont {T.~P.}\ \bibnamefont {Alegre}}, \bibinfo
  {author} {\bibfnamefont {M.}~\bibnamefont {Loncar}}, \ and\ \bibinfo {author}
  {\bibfnamefont {G.~S.}\ \bibnamefont {Wiederhecker}},\ }\bibfield  {title}
  {\bibinfo {title} {Cross-{Polarized} {Stimulated} {Brillouin} {Scattering} in
  {Lithium} {Niobate} {Waveguides}},\ }\href
  {\doibase10.1103/PhysRevLett.134.113601} {\bibfield  {journal} {\bibinfo
  {journal} {Phys. Rev. Lett.}\ }\textbf {\bibinfo {volume} {134}},\ \bibinfo
  {pages} {113601} (\bibinfo {year} {2025})}\BibitemShut {NoStop}%
\bibitem [{\citenamefont {Yang}\ \emph {,~et~al.}(2023)\citenamefont {Yang},
  \citenamefont {Wang}, \citenamefont {Zhu}, \citenamefont {Xu}, \citenamefont
  {Zhang}, \citenamefont {Lu}, \citenamefont {Zeng}, \citenamefont {Dong},
  \citenamefont {Sun}, \citenamefont {Guo},\ and\ \citenamefont
  {Zou}}]{yang_stimulated_2023}%
  \BibitemOpen
  \bibfield  {author} {\bibinfo {author} {\bibfnamefont {Y.-H.}\ \bibnamefont
  {Yang}}, \bibinfo {author} {\bibfnamefont {J.-Q.}\ \bibnamefont {Wang}},
  \bibinfo {author} {\bibfnamefont {Z.-X.}\ \bibnamefont {Zhu}}, \bibinfo
  {author} {\bibfnamefont {X.-B.}\ \bibnamefont {Xu}}, \bibinfo {author}
  {\bibfnamefont {Q.}~\bibnamefont {Zhang}}, \bibinfo {author} {\bibfnamefont
  {J.}~\bibnamefont {Lu}}, \bibinfo {author} {\bibfnamefont {Y.}~\bibnamefont
  {Zeng}}, \bibinfo {author} {\bibfnamefont {C.-H.}\ \bibnamefont {Dong}},
  \bibinfo {author} {\bibfnamefont {L.}~\bibnamefont {Sun}}, \bibinfo {author}
  {\bibfnamefont {G.-C.}\ \bibnamefont {Guo}}, \ and\ \bibinfo {author}
  {\bibfnamefont {C.-L.}\ \bibnamefont {Zou}},\ }\bibfield  {title} {\bibinfo
  {title} {Stimulated {Brillouin} interaction between guided phonons and
  photons in a lithium niobate waveguide},\ }\href
  {\doibase10.1007/s11433-023-2272-y} {\bibfield  {journal} {\bibinfo
  {journal} {Sci. China Phys. Mech.}\ }\textbf {\bibinfo {volume} {67}},\
  \bibinfo {pages} {214221} (\bibinfo {year} {2023})}\BibitemShut {NoStop}%
\bibitem [{\citenamefont {Yang}\ \emph {,~et~al.}(2025)\citenamefont {Yang},
  \citenamefont {Wang}, \citenamefont {Zhu}, \citenamefont {Zeng},
  \citenamefont {Li}, \citenamefont {Zhang}, \citenamefont {Lu}, \citenamefont
  {Zhang}, \citenamefont {Wang}, \citenamefont {Dong}, \citenamefont {Xu},
  \citenamefont {Guo}, \citenamefont {Sun},\ and\ \citenamefont
  {Zou}}]{yang2025multi}%
  \BibitemOpen
  \bibfield  {author} {\bibinfo {author} {\bibfnamefont {Y.-H.}\ \bibnamefont
  {Yang}}, \bibinfo {author} {\bibfnamefont {J.-Q.}\ \bibnamefont {Wang}},
  \bibinfo {author} {\bibfnamefont {Z.-X.}\ \bibnamefont {Zhu}}, \bibinfo
  {author} {\bibfnamefont {Y.}~\bibnamefont {Zeng}}, \bibinfo {author}
  {\bibfnamefont {M.}~\bibnamefont {Li}}, \bibinfo {author} {\bibfnamefont
  {Y.-L.}\ \bibnamefont {Zhang}}, \bibinfo {author} {\bibfnamefont
  {J.}~\bibnamefont {Lu}}, \bibinfo {author} {\bibfnamefont {Q.}~\bibnamefont
  {Zhang}}, \bibinfo {author} {\bibfnamefont {W.}~\bibnamefont {Wang}},
  \bibinfo {author} {\bibfnamefont {C.-H.}\ \bibnamefont {Dong}},~et~al.,\
  }\bibfield  {title} {\bibinfo {title} {Multi-channel microwave-to-optics
  conversion utilizing a hybrid photonic-phononic waveguide},\ }\href
  {https://arxiv.org/abs/2509.10052} {\bibfield  {journal} {\bibinfo  {journal}
  {arXiv preprint:2509.10052}\ } (\bibinfo {year} {2025})}\BibitemShut
  {NoStop}%
\bibitem [{\citenamefont {Zhang}\ \emph {,~et~al.}(2017)\citenamefont {Zhang},
  \citenamefont {Dong}, \citenamefont {Zou}, \citenamefont {Zou}, \citenamefont
  {Wang},\ and\ \citenamefont {Guo}}]{zhang_optomechanical_2017}%
  \BibitemOpen
  \bibfield  {author} {\bibinfo {author} {\bibfnamefont {Y.-L.}\ \bibnamefont
  {Zhang}}, \bibinfo {author} {\bibfnamefont {C.-H.}\ \bibnamefont {Dong}},
  \bibinfo {author} {\bibfnamefont {C.-L.}\ \bibnamefont {Zou}}, \bibinfo
  {author} {\bibfnamefont {X.-B.}\ \bibnamefont {Zou}}, \bibinfo {author}
  {\bibfnamefont {Y.-D.}\ \bibnamefont {Wang}}, \ and\ \bibinfo {author}
  {\bibfnamefont {G.-C.}\ \bibnamefont {Guo}},\ }\bibfield  {title} {\bibinfo
  {title} {Optomechanical devices based on traveling-wave microresonators},\
  }\href {\doibase10.1103/PhysRevA.95.043815} {\bibfield  {journal} {\bibinfo
  {journal} {Phys. Rev. A}\ }\textbf {\bibinfo {volume} {95}},\ \bibinfo
  {pages} {043815} (\bibinfo {year} {2017})}\BibitemShut {NoStop}%
\bibitem [{\citenamefont {Yang}\ \emph
  {,~et~al.}(2024{\natexlab{a}})\citenamefont {Yang}, \citenamefont {Wang},
  \citenamefont {Xu}, \citenamefont {Li}, \citenamefont {Zhang}, \citenamefont
  {Pan}, \citenamefont {Xiao}, \citenamefont {Wang}, \citenamefont {Guo},
  \citenamefont {Sun},\ and\ \citenamefont {Zou}}]{yang_proposal_2024}%
  \BibitemOpen
  \bibfield  {author} {\bibinfo {author} {\bibfnamefont {Y.-H.}\ \bibnamefont
  {Yang}}, \bibinfo {author} {\bibfnamefont {J.-Q.}\ \bibnamefont {Wang}},
  \bibinfo {author} {\bibfnamefont {X.-B.}\ \bibnamefont {Xu}}, \bibinfo
  {author} {\bibfnamefont {M.}~\bibnamefont {Li}}, \bibinfo {author}
  {\bibfnamefont {Y.-L.}\ \bibnamefont {Zhang}}, \bibinfo {author}
  {\bibfnamefont {X.}~\bibnamefont {Pan}}, \bibinfo {author} {\bibfnamefont
  {L.}~\bibnamefont {Xiao}}, \bibinfo {author} {\bibfnamefont {W.}~\bibnamefont
  {Wang}}, \bibinfo {author} {\bibfnamefont {G.-c.}\ \bibnamefont {Guo}},
  \bibinfo {author} {\bibfnamefont {L.}~\bibnamefont {Sun}}, \ and\ \bibinfo
  {author} {\bibfnamefont {C.-l.}\ \bibnamefont {Zou}},\ }\bibfield  {title}
  {\bibinfo {title} {Proposal for {Brillouin} microwave-to-optical conversion
  on a chip [{Invited}]},\ }\href {\doibase10.1364/OME.534817} {\bibfield
  {journal} {\bibinfo  {journal} {Opt. Mater. Express}\ }\textbf {\bibinfo
  {volume} {14}},\ \bibinfo {pages} {2400} (\bibinfo {year}
  {2024}{\natexlab{a}})}\BibitemShut {NoStop}%
\bibitem [{\citenamefont {Shao}\ \emph {,~et~al.}(2022)\citenamefont {Shao},
  \citenamefont {Zhu}, \citenamefont {Colangelo}, \citenamefont {Lee},
  \citenamefont {Sinclair}, \citenamefont {Hu}, \citenamefont {Rakich},
  \citenamefont {Lai}, \citenamefont {Berggren},\ and\ \citenamefont
  {Lon{\v{c}}ar}}]{Shao2022}%
  \BibitemOpen
  \bibfield  {author} {\bibinfo {author} {\bibfnamefont {L.}~\bibnamefont
  {Shao}}, \bibinfo {author} {\bibfnamefont {D.}~\bibnamefont {Zhu}}, \bibinfo
  {author} {\bibfnamefont {M.}~\bibnamefont {Colangelo}}, \bibinfo {author}
  {\bibfnamefont {D.}~\bibnamefont {Lee}}, \bibinfo {author} {\bibfnamefont
  {N.}~\bibnamefont {Sinclair}}, \bibinfo {author} {\bibfnamefont
  {Y.}~\bibnamefont {Hu}}, \bibinfo {author} {\bibfnamefont {P.~T.}\
  \bibnamefont {Rakich}}, \bibinfo {author} {\bibfnamefont {K.}~\bibnamefont
  {Lai}}, \bibinfo {author} {\bibfnamefont {K.~K.}\ \bibnamefont {Berggren}}, \
  and\ \bibinfo {author} {\bibfnamefont {M.}~\bibnamefont {Lon{\v{c}}ar}},\
  }\bibfield  {title} {\bibinfo {title} {{Electrical control of surface
  acoustic waves}},\ }\href {\doibase10.1038/s41928-022-00773-3} {\bibfield
  {journal} {\bibinfo  {journal} {Nature Electronics}\ }\textbf {\bibinfo
  {volume} {5}},\ \bibinfo {pages} {348} (\bibinfo {year} {2022})}\BibitemShut
  {NoStop}%
\bibitem [{\citenamefont {Hann}\ \emph {,~et~al.}(2019)\citenamefont {Hann},
  \citenamefont {Zou}, \citenamefont {Zhang}, \citenamefont {Chu},
  \citenamefont {Schoelkopf}, \citenamefont {Girvin},\ and\ \citenamefont
  {Jiang}}]{Hann2019}%
  \BibitemOpen
  \bibfield  {author} {\bibinfo {author} {\bibfnamefont {C.~T.}\ \bibnamefont
  {Hann}}, \bibinfo {author} {\bibfnamefont {C.-L.}\ \bibnamefont {Zou}},
  \bibinfo {author} {\bibfnamefont {Y.}~\bibnamefont {Zhang}}, \bibinfo
  {author} {\bibfnamefont {Y.}~\bibnamefont {Chu}}, \bibinfo {author}
  {\bibfnamefont {R.~J.}\ \bibnamefont {Schoelkopf}}, \bibinfo {author}
  {\bibfnamefont {S.~M.}\ \bibnamefont {Girvin}}, \ and\ \bibinfo {author}
  {\bibfnamefont {L.}~\bibnamefont {Jiang}},\ }\bibfield  {title} {\bibinfo
  {title} {{Hardware-Efficient Quantum Random Access Memory with Hybrid Quantum
  Acoustic Systems}},\ }\href {\doibase10.1103/PhysRevLett.123.250501}
  {\bibfield  {journal} {\bibinfo  {journal} {Phys. Rev. Lett.}\ }\textbf
  {\bibinfo {volume} {123}},\ \bibinfo {pages} {250501} (\bibinfo {year}
  {2019})}\BibitemShut {NoStop}%
\bibitem [{\citenamefont {Yang}\ \emph
  {,~et~al.}(2024{\natexlab{b}})\citenamefont {Yang}, \citenamefont
  {Kladari{\'{c}}}, \citenamefont {Drimmer}, \citenamefont {von L{\"{u}}pke},
  \citenamefont {Lenterman}, \citenamefont {Bus}, \citenamefont {Marti},
  \citenamefont {Fadel},\ and\ \citenamefont {Chu}}]{Yang2024}%
  \BibitemOpen
  \bibfield  {author} {\bibinfo {author} {\bibfnamefont {Y.}~\bibnamefont
  {Yang}}, \bibinfo {author} {\bibfnamefont {I.}~\bibnamefont
  {Kladari{\'{c}}}}, \bibinfo {author} {\bibfnamefont {M.}~\bibnamefont
  {Drimmer}}, \bibinfo {author} {\bibfnamefont {U.}~\bibnamefont {von
  L{\"{u}}pke}}, \bibinfo {author} {\bibfnamefont {D.}~\bibnamefont
  {Lenterman}}, \bibinfo {author} {\bibfnamefont {J.}~\bibnamefont {Bus}},
  \bibinfo {author} {\bibfnamefont {S.}~\bibnamefont {Marti}}, \bibinfo
  {author} {\bibfnamefont {M.}~\bibnamefont {Fadel}}, \ and\ \bibinfo {author}
  {\bibfnamefont {Y.}~\bibnamefont {Chu}},\ }\bibfield  {title} {\bibinfo
  {title} {{A mechanical qubit}},\ }\href {\doibase10.1126/science.adr2464}
  {\bibfield  {journal} {\bibinfo  {journal} {Science}\ }\textbf {\bibinfo
  {volume} {386}},\ \bibinfo {pages} {783} (\bibinfo {year}
  {2024}{\natexlab{b}})}\BibitemShut {NoStop}%
\bibitem [{\citenamefont {Xu}\ \emph
  {,~et~al.}(2025{\natexlab{a}})\citenamefont {Xu}, \citenamefont {Xiao},
  \citenamefont {Zhang}, \citenamefont {Wang}, \citenamefont {Wang},
  \citenamefont {Zeng}, \citenamefont {Yang}, \citenamefont {Wang},
  \citenamefont {Pan}, \citenamefont {Guo}, \citenamefont {Sun},\ and\
  \citenamefont {Zou}}]{Xu2025QAD}%
  \BibitemOpen
  \bibfield  {author} {\bibinfo {author} {\bibfnamefont {X.-B.}\ \bibnamefont
  {Xu}}, \bibinfo {author} {\bibfnamefont {L.}~\bibnamefont {Xiao}}, \bibinfo
  {author} {\bibfnamefont {B.}~\bibnamefont {Zhang}}, \bibinfo {author}
  {\bibfnamefont {W.}~\bibnamefont {Wang}}, \bibinfo {author} {\bibfnamefont
  {J.-Q.}\ \bibnamefont {Wang}}, \bibinfo {author} {\bibfnamefont
  {Y.}~\bibnamefont {Zeng}}, \bibinfo {author} {\bibfnamefont {Y.-H.}\
  \bibnamefont {Yang}}, \bibinfo {author} {\bibfnamefont {B.-Z.}\ \bibnamefont
  {Wang}}, \bibinfo {author} {\bibfnamefont {X.}~\bibnamefont {Pan}}, \bibinfo
  {author} {\bibfnamefont {G.-C.}\ \bibnamefont {Guo}},~et~al.,\ }\bibfield
  {title} {\bibinfo {title} {{Proposal of cavity quantum acoustodynamics
  platform based on Lithium Niobate-on-Sapphire chip}},\ }\href
  {http://arxiv.org/abs/2509.14728} {\bibfield  {journal} {\bibinfo  {journal}
  {arXiv preprint: 2509.14728}\ } (\bibinfo {year}
  {2025}{\natexlab{a}})}\BibitemShut {NoStop}%
\bibitem [{\citenamefont {Xu}\ \emph
  {,~et~al.}(2025{\natexlab{b}})\citenamefont {Xu}, \citenamefont {Zhu},
  \citenamefont {Yang}, \citenamefont {Wang}, \citenamefont {Zeng},
  \citenamefont {Zou}, \citenamefont {Lu}, \citenamefont {Zhang}, \citenamefont
  {Wang}, \citenamefont {Guo}, \citenamefont {Sun},\ and\ \citenamefont
  {Zou}}]{xu2025magnetic}%
  \BibitemOpen
  \bibfield  {author} {\bibinfo {author} {\bibfnamefont {X.-B.}\ \bibnamefont
  {Xu}}, \bibinfo {author} {\bibfnamefont {Z.-X.}\ \bibnamefont {Zhu}},
  \bibinfo {author} {\bibfnamefont {Y.-H.}\ \bibnamefont {Yang}}, \bibinfo
  {author} {\bibfnamefont {J.-Q.}\ \bibnamefont {Wang}}, \bibinfo {author}
  {\bibfnamefont {Y.}~\bibnamefont {Zeng}}, \bibinfo {author} {\bibfnamefont
  {J.-H.}\ \bibnamefont {Zou}}, \bibinfo {author} {\bibfnamefont
  {J.}~\bibnamefont {Lu}}, \bibinfo {author} {\bibfnamefont {Y.-L.}\
  \bibnamefont {Zhang}}, \bibinfo {author} {\bibfnamefont {W.}~\bibnamefont
  {Wang}}, \bibinfo {author} {\bibfnamefont {G.-C.}\ \bibnamefont
  {Guo}},~et~al.,\ }\bibfield  {title} {\bibinfo {title} {Magnetic-free optical
  mode degeneracy lifting in lithium niobate microring resonators},\ }\href
  {https://arxiv.org/abs/2509.01940} {\bibfield  {journal} {\bibinfo  {journal}
  {arXiv preprint:2509.10052}\ } (\bibinfo {year}
  {2025}{\natexlab{b}})}\BibitemShut {NoStop}%
\bibitem [{sm()}]{sm}%
  \BibitemOpen
  \bibinfo {howpublished} {See the Supplemental Materials for
  details about the device, fabrications, and theoretical
  derivations.}\BibitemShut {Stop}%
\bibitem [{\citenamefont {Sipe}\ and\ \citenamefont
  {Steel}(2016)}]{sipe_hamiltonian_2016}%
  \BibitemOpen
  \bibfield  {author} {\bibinfo {author} {\bibfnamefont {J.~E.}\ \bibnamefont
  {Sipe}}\ and\ \bibinfo {author} {\bibfnamefont {M.~J.}\ \bibnamefont
  {Steel}},\ }\bibfield  {title} {\bibinfo {title} {A {Hamiltonian} treatment
  of stimulated {Brillouin} scattering in nanoscale integrated waveguides},\
  }\href {\doibase10.1088/1367-2630/18/4/045004} {\bibfield  {journal}
  {\bibinfo  {journal} {New J. Phys.}\ }\textbf {\bibinfo {volume} {18}},\
  \bibinfo {pages} {045004} (\bibinfo {year} {2016})}\BibitemShut {NoStop}%
\bibitem [{\citenamefont {Chen}\ \emph {,~et~al.}(2025)\citenamefont {Chen},
  \citenamefont {Wang}, \citenamefont {Yang}, \citenamefont {Zhu},
  \citenamefont {Xu}, \citenamefont {Li}, \citenamefont {Ren}, \citenamefont
  {Guo},\ and\ \citenamefont {Zou}}]{chen2025fiber}%
  \BibitemOpen
  \bibfield  {author} {\bibinfo {author} {\bibfnamefont {X.}~\bibnamefont
  {Chen}}, \bibinfo {author} {\bibfnamefont {J.-Q.}\ \bibnamefont {Wang}},
  \bibinfo {author} {\bibfnamefont {Y.-H.}\ \bibnamefont {Yang}}, \bibinfo
  {author} {\bibfnamefont {Z.-X.}\ \bibnamefont {Zhu}}, \bibinfo {author}
  {\bibfnamefont {X.-B.}\ \bibnamefont {Xu}}, \bibinfo {author} {\bibfnamefont
  {M.}~\bibnamefont {Li}}, \bibinfo {author} {\bibfnamefont {X.-F.}\
  \bibnamefont {Ren}}, \bibinfo {author} {\bibfnamefont {G.-C.}\ \bibnamefont
  {Guo}}, \ and\ \bibinfo {author} {\bibfnamefont {C.-L.}\ \bibnamefont
  {Zou}},\ }\bibfield  {title} {\bibinfo {title} {Fiber-to-chip grating
  couplers for lithium niobate on sapphire},\ }\href
  {https://arxiv.org/abs/2510.02089} {\bibfield  {journal} {\bibinfo  {journal}
  {arXiv preprint:2510.02089}\ } (\bibinfo {year} {2025})}\BibitemShut
  {NoStop}%
\bibitem [{\citenamefont {Shen}\ \emph {,~et~al.}(2018)\citenamefont {Shen},
  \citenamefont {Zhang}, \citenamefont {Chen}, \citenamefont {Sun},
  \citenamefont {Zou}, \citenamefont {Guo}, \citenamefont {Zou},\ and\
  \citenamefont {Dong}}]{Shen2018}%
  \BibitemOpen
  \bibfield  {author} {\bibinfo {author} {\bibfnamefont {Z.}~\bibnamefont
  {Shen}}, \bibinfo {author} {\bibfnamefont {Y.-L.}\ \bibnamefont {Zhang}},
  \bibinfo {author} {\bibfnamefont {Y.}~\bibnamefont {Chen}}, \bibinfo {author}
  {\bibfnamefont {F.-W.}\ \bibnamefont {Sun}}, \bibinfo {author} {\bibfnamefont
  {X.-B.}\ \bibnamefont {Zou}}, \bibinfo {author} {\bibfnamefont {G.-C.}\
  \bibnamefont {Guo}}, \bibinfo {author} {\bibfnamefont {C.-L.}\ \bibnamefont
  {Zou}}, \ and\ \bibinfo {author} {\bibfnamefont {C.-H.}\ \bibnamefont
  {Dong}},\ }\bibfield  {title} {\bibinfo {title} {{Reconfigurable
  optomechanical circulator and directional amplifier}},\ }\href
  {\doibase10.1038/s41467-018-04187-8} {\bibfield  {journal} {\bibinfo
  {journal} {Nat. Commun.}\ }\textbf {\bibinfo {volume} {9}},\ \bibinfo {pages}
  {1797} (\bibinfo {year} {2018})}\BibitemShut {NoStop}%
\bibitem [{\citenamefont {Grudinin}\ \emph {,~et~al.}(2009)\citenamefont
  {Grudinin}, \citenamefont {Matsko},\ and\ \citenamefont
  {Maleki}}]{Grudinin2009}%
  \BibitemOpen
  \bibfield  {author} {\bibinfo {author} {\bibfnamefont {I.~S.}\ \bibnamefont
  {Grudinin}}, \bibinfo {author} {\bibfnamefont {A.~B.}\ \bibnamefont
  {Matsko}}, \ and\ \bibinfo {author} {\bibfnamefont {L.}~\bibnamefont
  {Maleki}},\ }\bibfield  {title} {\bibinfo {title} {{Brillouin Lasing with a
  CaF2 Whispering Gallery Mode Resonator}},\ }\href
  {\doibase10.1103/PhysRevLett.102.043902} {\bibfield  {journal} {\bibinfo
  {journal} {Phys. Rev. Lett.}\ }\textbf {\bibinfo {volume} {102}},\ \bibinfo
  {pages} {043902} (\bibinfo {year} {2009})}\BibitemShut {NoStop}%
\bibitem [{\citenamefont {Lee}\ \emph {,~et~al.}(2012)\citenamefont {Lee},
  \citenamefont {Chen}, \citenamefont {Li}, \citenamefont {Yang}, \citenamefont
  {Jeon}, \citenamefont {Painter},\ and\ \citenamefont {Vahala}}]{Lee2012}%
  \BibitemOpen
  \bibfield  {author} {\bibinfo {author} {\bibfnamefont {H.}~\bibnamefont
  {Lee}}, \bibinfo {author} {\bibfnamefont {T.}~\bibnamefont {Chen}}, \bibinfo
  {author} {\bibfnamefont {J.}~\bibnamefont {Li}}, \bibinfo {author}
  {\bibfnamefont {K.~Y.}\ \bibnamefont {Yang}}, \bibinfo {author}
  {\bibfnamefont {S.}~\bibnamefont {Jeon}}, \bibinfo {author} {\bibfnamefont
  {O.}~\bibnamefont {Painter}}, \ and\ \bibinfo {author} {\bibfnamefont
  {K.~J.}\ \bibnamefont {Vahala}},\ }\bibfield  {title} {\bibinfo {title}
  {{Chemically etched ultrahigh-Q wedge-resonator on a silicon chip}},\ }\href
  {\doibase10.1038/nphoton.2012.109} {\bibfield  {journal} {\bibinfo  {journal}
  {Nat. Photon.}\ }\textbf {\bibinfo {volume} {6}},\ \bibinfo {pages} {369}
  (\bibinfo {year} {2012})}\BibitemShut {NoStop}%
\bibitem [{\citenamefont {Cohen}\ \emph {,~et~al.}(2015)\citenamefont {Cohen},
  \citenamefont {Meenehan}, \citenamefont {MacCabe}, \citenamefont
  {Groblacher}, \citenamefont {Safavi-Naeini}, \citenamefont {Marsili},
  \citenamefont {Shaw},\ and\ \citenamefont {Painter}}]{cohen_phonon_2015}%
  \BibitemOpen
  \bibfield  {author} {\bibinfo {author} {\bibfnamefont {J.~D.}\ \bibnamefont
  {Cohen}}, \bibinfo {author} {\bibfnamefont {S.~M.}\ \bibnamefont {Meenehan}},
  \bibinfo {author} {\bibfnamefont {G.~S.}\ \bibnamefont {MacCabe}}, \bibinfo
  {author} {\bibfnamefont {S.}~\bibnamefont {Groblacher}}, \bibinfo {author}
  {\bibfnamefont {A.~H.}\ \bibnamefont {Safavi-Naeini}}, \bibinfo {author}
  {\bibfnamefont {F.}~\bibnamefont {Marsili}}, \bibinfo {author} {\bibfnamefont
  {M.~D.}\ \bibnamefont {Shaw}}, \ and\ \bibinfo {author} {\bibfnamefont
  {O.}~\bibnamefont {Painter}},\ }\bibfield  {title} {\bibinfo {title} {Phonon
  counting and intensity interferometry of a nanomechanical resonator},\ }\href
  {\doibase10.1038/nature14349} {\bibfield  {journal} {\bibinfo  {journal}
  {Nature}\ }\textbf {\bibinfo {volume} {520}},\ \bibinfo {pages} {522}
  (\bibinfo {year} {2015})}\BibitemShut {NoStop}%
\bibitem [{\citenamefont {Xu}\ \emph
  {,~et~al.}(2022{\natexlab{b}})\citenamefont {Xu}, \citenamefont {Wang},
  \citenamefont {Yang}, \citenamefont {Wang}, \citenamefont {Zhang},
  \citenamefont {Wang}, \citenamefont {Dong}, \citenamefont {Sun},
  \citenamefont {Guo},\ and\ \citenamefont {Zou}}]{xu_high-frequency_2022}%
  \BibitemOpen
  \bibfield  {author} {\bibinfo {author} {\bibfnamefont {X.-B.}\ \bibnamefont
  {Xu}}, \bibinfo {author} {\bibfnamefont {J.-Q.}\ \bibnamefont {Wang}},
  \bibinfo {author} {\bibfnamefont {Y.-H.}\ \bibnamefont {Yang}}, \bibinfo
  {author} {\bibfnamefont {W.}~\bibnamefont {Wang}}, \bibinfo {author}
  {\bibfnamefont {Y.-L.}\ \bibnamefont {Zhang}}, \bibinfo {author}
  {\bibfnamefont {B.-Z.}\ \bibnamefont {Wang}}, \bibinfo {author}
  {\bibfnamefont {C.-H.}\ \bibnamefont {Dong}}, \bibinfo {author}
  {\bibfnamefont {L.}~\bibnamefont {Sun}}, \bibinfo {author} {\bibfnamefont
  {G.-C.}\ \bibnamefont {Guo}}, \ and\ \bibinfo {author} {\bibfnamefont
  {C.-L.}\ \bibnamefont {Zou}},\ }\bibfield  {title} {\bibinfo {title}
  {High-frequency traveling-wave phononic cavity with sub-micron wavelength},\
  }\href {\doibase10.1063/5.0086751} {\bibfield  {journal} {\bibinfo  {journal}
  {App. Phys. Lett.}\ }\textbf {\bibinfo {volume} {120}},\ \bibinfo {pages}
  {163503} (\bibinfo {year} {2022}{\natexlab{b}})}\BibitemShut {NoStop}%
\bibitem [{\citenamefont {Manenti}\ \emph {,~et~al.}(2017)\citenamefont
  {Manenti}, \citenamefont {Kockum}, \citenamefont {Patterson}, \citenamefont
  {Behrle}, \citenamefont {Rahamim}, \citenamefont {Tancredi}, \citenamefont
  {Nori},\ and\ \citenamefont {Leek}}]{manenti_circuit_2017}%
  \BibitemOpen
  \bibfield  {author} {\bibinfo {author} {\bibfnamefont {R.}~\bibnamefont
  {Manenti}}, \bibinfo {author} {\bibfnamefont {A.~F.}\ \bibnamefont {Kockum}},
  \bibinfo {author} {\bibfnamefont {A.}~\bibnamefont {Patterson}}, \bibinfo
  {author} {\bibfnamefont {T.}~\bibnamefont {Behrle}}, \bibinfo {author}
  {\bibfnamefont {J.}~\bibnamefont {Rahamim}}, \bibinfo {author} {\bibfnamefont
  {G.}~\bibnamefont {Tancredi}}, \bibinfo {author} {\bibfnamefont
  {F.}~\bibnamefont {Nori}}, \ and\ \bibinfo {author} {\bibfnamefont {P.~J.}\
  \bibnamefont {Leek}},\ }\bibfield  {title} {\bibinfo {title} {Circuit quantum
  acoustodynamics with surface acoustic waves},\ }\href
  {\doibase10.1038/s41467-017-01063-9} {\bibfield  {journal} {\bibinfo
  {journal} {Nat. Commun.}\ }\textbf {\bibinfo {volume} {8}},\ \bibinfo {pages}
  {975} (\bibinfo {year} {2017})}\BibitemShut {NoStop}%
\end{thebibliography}
\end{document}